\documentclass[a4paper,twocolumn,english,aps,prl,superscriptaddress]{revtex4-1}
\usepackage[T1]{fontenc}
\usepackage[latin9]{inputenc}
\usepackage{color}
\usepackage{babel}
\usepackage{amsmath}
\usepackage{amssymb}
\usepackage{graphicx}
\DeclareGraphicsExtensions{.pdf}
\usepackage{amsmath,amssymb,bbold,bm,color}
\usepackage{float}
\usepackage{epstopdf}
\usepackage{tabularx}
\usepackage{braket}
\usepackage{booktabs}
\usepackage{array}
\usepackage{blindtext}
\usepackage{natbib}
\usepackage[unicode=true,pdfusetitle,
 bookmarks=true,bookmarksnumbered=false,bookmarksopen=false,
 breaklinks=false,pdfborder={0 0 1},backref=false,colorlinks=false]
 {hyperref}

\usepackage{ulem} 

\makeatletter

\usepackage{babel}
\usepackage{babel}

\begin{document}

\title{Geometric Superfluid Weight in Quasicrystals}

\author{Junsong Sun}
\affiliation{Department of Physics and Astronomy, Seoul National University, Seoul 08826, Korea}
\affiliation{School of Physics, Beihang University,
Beijing, 100191, China}

\author{Huaiming Guo}
\email{hmguo@buaa.edu.cn}
\affiliation{School of Physics, Beihang University,
Beijing, 100191, China}

\author{Bohm-Jung Yang}
\email{bjyang@snu.ac.kr}
\affiliation{Department of Physics and Astronomy, Seoul National University, Seoul 08826, Korea}
\affiliation{Center for Theoretical Physics (CTP), Seoul National University, Seoul 08826, Korea}
\affiliation{Institute of Applied Physics, Seoul National University, Seoul 08826, Korea}

\begin{abstract}
We study the geometric contribution to the superfluidity in quasicrystals in which the conventional momentum-space quantum geometric tensor cannot be defined due to the lack of translational invariance.
Based on the correspondence between the momentum and magnetic flux, we propose a general framework to define the quantum metric in flux space and reveal its contribution to the superfluid
weight in quasicrystalline superconductors. As an application, we investigate the attractive Hubbard model on various extrinsic and intrinsic quasicrystals with flat-energy spectra. In the weak-coupling limit, we reveal the geometric origin of the superfluid weight in quasicrystals with flat-energy spectra, and establish its connection to the quantum metric in flux-space. Moreover, by analyzing the spread of Wannier functions, we propose a general fluctuation mechanism that explains how quasiperiodicity modulates the integrated flux-space  quantum metric. Our theory provides a general way to examine the effect of the quantum geometry in systems lacking translational symmetry.
\end{abstract}

\date{\today}

\maketitle
\textit{\textcolor{blue}{Introduction.--}}
Quantum geometry has become a pivotal theoretical concept in understanding various fundamental physical phenomena in condensed matter physics. 
The geometry of a quantum state in a parameter space is fully encoded in the quantum geometric tensor (QGT)~\cite{Resta2011, Torma2022, PhysRevLett.131.240001, liutianyu2024,yu2025quantumgeometryquantummaterials}.
Its imaginary part is the well known Berry curvature, which has become an essential concept in the field of the topological phases of matter~\cite{RevModPhys.82.1959,RevModPhys.82.3045,RevModPhys.83.1057,RevModPhys.90.015001}. The real part of the QGT represents the quantum metric (QM), which measures the distance between neighboring quantum states. Recently, it has started to garner attention for its central role in explaining an increasingly diverse range of geometric phenomena in solids~\cite{gao2023quantum,wang2023quantum,tian2023evidence,rhim2020quantum,hwang2021geometric,PhysRevB.109.035134,kang2025measurements,doi:10.1126/science.ado6049,Balazs2023}. Specifically, in the context of flat-band superconductivity, while conventional BCS theory fails to account for the finite supercurrent in a flat band, extensive studies have established a connection between the superfluid weight and QM in many flat-band superconducting systems~\cite{PhysRevB.105.024502, PhysRevB.96.064511, PhysRevB.98.134513, PhysRevB.94.245149, PhysRevB.106.014518, PhysRevB.95.024515, PhysRevB.109.214518, PhysRevA.103.043329, peotta2023, PhysRevLett.128.087002,PhysRevLett.123.237002, PhysRevLett.124.167002,PhysRevB.106.014518}. 

The QM of the electrons in solids has primarily been parametrized by the Bloch momentum. However, there are various quasicrystalline superconducting systems~\cite{PhysRevB.94.125408,PhysRevB.93.104504,Kamiya2018,PhysRevB.100.014510,PhysRevB.102.115108,PhysRevResearch.3.023195,PhysRevResearch.5.043164,Uri2023,PhysRevLett.133.136002,PhysRevB.110.134508,Tokumoto2024,PhysRevB.109.134504,Ghadimi2025,PhysRevLett.56.2740(1986),PhysRevB.100.014510,PhysRevB.102.064210(2020),zhang2024FBQC,Diaz-Reynoso_2024,MannaSourav2024,DebikaDebnath2025}, some of which exhibit nearly degenerate energy spectra similar to those in flat-band crystals\cite{PhysRevB.100.014510,zhang2024FBQC,Diaz-Reynoso_2024,PhysRevLett.56.2740(1986),PhysRevB.102.064210(2020)}. 
Given the lack of the translational symmetry, one might wonder whether QM still plays a key role in such systems where QM theory for Bloch electronic states no longer applies. The challenge lies in suitably defining QM using a real-space approach, analogous to its definition in momentum-space for periodic systems. To date, this issue remains open and largely unexplored~\cite{Resta2018,PhysRevLett.122.166602,PhysRevB.107.205133,PhysRevB.111.134201,marsal2025QM_quasicrystal}.


In this work, we develop a general real-space framework for characterize quantum geometry in aperiodic systems lacking translational symmetry and reveal the geometric origin of superfluidity in quasicrystalline superconductors. In the weak-coupling limit, we universally establish a direct relation between the superfluid weight and a flux-space QM defined via flux insertion. Applying this general framework to the attractive Hubbard model on various extrinsic and intrinsic quasicrystals (QCs) that host a flat-energy spectra, we demonstrate how modulation of the quantum metric controls the superfluid weight.  Furthermore, we show that quasiperiodicity tunes the integrated flux-space quantum metric by inducing fluctuations in the spread of Wannier functions.

\begin{figure}[hbpt]
  \includegraphics[width=8.8cm]{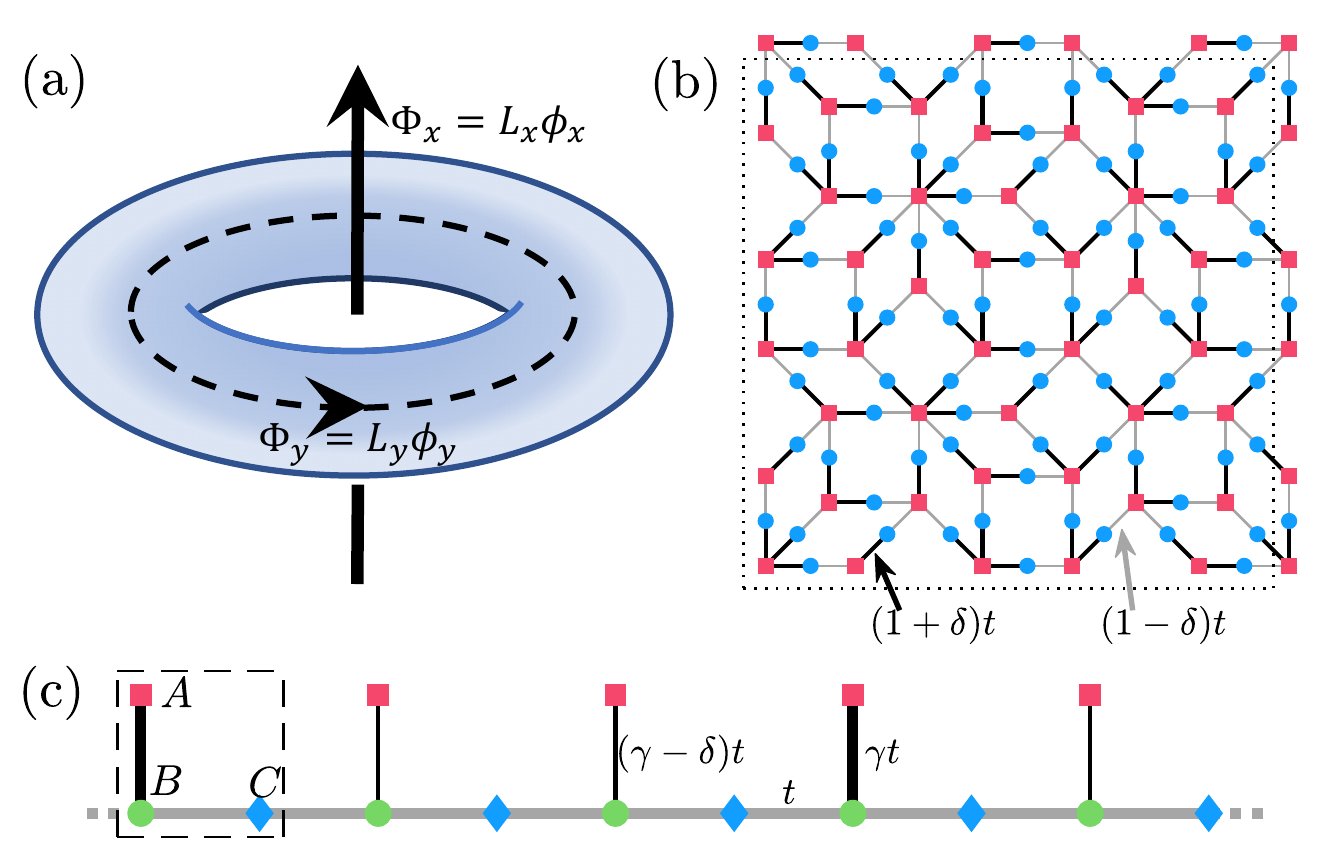}
  \caption{
(a) Illustration of magnetic fluxes threading a 2D torus (reducing to a ring in the 1D case), used to formulate the flux-space quantum metric. (b) Structure of the 2D Amman-Beenker (AB)-QC composed of two sublattices, featuring two hopping amplitudes, \( (1+\delta)t \) and \( (1-\delta)t \). (c) Schematic of the 1D Stub-QC, which consists of three sites in each unit cell and becomes quasiperiodic by modulating the vertical hopping amplitudes according to the Fibonacci sequence. 
}\label{fig1}
\end{figure}
\textit{\textcolor{blue}{Flux-space formulation of quantum metric.--}}
In a general $d$-dimensional periodic crystal, the quantum distance $D$ between two infinitesimally close quantum states is defined through
\begin{equation*}
D^2 = 1 - \bigl|\langle u_{s,\boldsymbol{k}}|u_{s,\boldsymbol{k}+d\boldsymbol{k}}\rangle\bigr|^2
\approx \tfrac{1}{2} \sum_{\alpha,\beta} g_{s,\alpha\beta}(\boldsymbol{k})dk_\alpha dk_\beta,
\end{equation*}
which gives the momentum-space QM $g_{s,\alpha\beta}(\boldsymbol{k})$~\cite{Provost1980,liutianyu2024}. Formally,
\begin{align}
g_{s,\alpha\beta}(\boldsymbol{k})\equiv \left.\frac{d^2}{dq_\alpha dq_\beta}[1-|\langle u_s(\boldsymbol{k})|u_s(\boldsymbol{k}+\boldsymbol{q})\rangle|^2]\right|_{q_{\alpha,\beta}=0},
\end{align}
where ${q}_\alpha$ denotes the $\alpha$-th component of the momentum displacement $\boldsymbol{q}$ and $s$ labels the band index.
While this definition is well established for translationally invariant systems, it cannot be directly applied to aperiodic systems, where the Bloch states are no longer well-defined.
In this situation, it is worth noting that a constant vector potential $\boldsymbol{\phi}$ introduced to the Hamiltonian through the Peierls substitution, $t_{ij} \rightarrow t_{ij} e^{i\boldsymbol{\phi}\cdot\boldsymbol{r}_{ij}}$, plays a role analogous to that of $\boldsymbol{q}$. Consequently, replacing $q_\alpha$ by $\phi_\alpha$ yields the same QM, but the physical meaning changes to quantify the quantum distance between the states $|u_s(\boldsymbol{k})\rangle$ and $|u_s(\boldsymbol{k}+\boldsymbol{\phi})\rangle$. 
In one-dimensional (1D) systems, $\boldsymbol{\phi}=\phi_x$ is a scalar corresponding to the magnetic flux thread through the closed chain. On the other hand, in two-dimensional (2D) systems, the vector $\boldsymbol{\phi}=(\phi_x,\phi_y)$ is
associated with the magnetic fluxes penetrating the two non-contractible loops of the system with a torus geometry along the $x$ and $y$ directions, respectively [see Fig.~\ref{fig1}(a)].


Interestingly, this viewpoint leads to a flux-space formulation of the QM, which can be directly applied to real-space Hamiltonians without translational symmetry. Specifically, for a general $d$-dimensional real-space Hamiltonian $\mathcal{H}^0_{\rm rs}(0)$, we introduce a flux $\boldsymbol{\phi}$ through the Peierls substitution and construct $\mathcal{H}^0_{\rm rs}(\boldsymbol{\phi})$. We then define a quantum distance matrix $M^{\mathrm{rs}}_s(\boldsymbol{\phi})$ that captures the distance between eigenvector subspaces $P_s(0)$ and $P_s(\boldsymbol{\phi})$ as
\begin{align}\label{eqMrs}
M^{\mathrm{rs}}_s(\boldsymbol{\phi}) = I - P^\dagger_s(0) P_s(\boldsymbol{\phi}) P^\dagger_s(\boldsymbol{\phi}) P_s(0),
\end{align}
where $I$ is an identity matrix, $P_s(\boldsymbol{\phi})=(v_1,v_2,\cdots,v_{N_s})$ is the matrix of eigenvectors corresponding to a chosen subset of eigenstates labeled by the subscript $s$ with \(v_i\) the \(i\)-th eigenvector within the subset $s$, and $N_s$ is the number of eigenstates in the $s$ sector. When $P_s(0)$ and $P_s(\boldsymbol{\phi})$ coincide exactly, the distance matrix reduces to the zero matrix, indicating zero distance. In contrast, when $P_s(0)$ and $P_s(\boldsymbol{\phi})$ are completely orthogonal, $M^{\mathrm{rs}}_s(\boldsymbol{\phi})$ becomes the identity matrix, signifying the maximal possible distance between them. Accordingly, a flux-space QM is defined as 
 \begin{align}\label{eqgni}
g_{s,\alpha\beta,i}^{\rm FS}=\left.\frac{d^2}{d\phi_\alpha d\phi_\beta}d^{\rm rs}_{s,i}\right|_{\phi_{\alpha,\beta}=0},
\end{align}
where \( d^{\rm rs}_{s,i}(\boldsymbol{\phi})\) denotes the \( i \)-th eigenvalue of $M^{\rm rs}_s(\boldsymbol{\phi})$.

Notably, in a finite-size system under periodic boundary condition, the proposed flux-space QM in real-space representation is naturally reduced to 
its momentum-space counterpart, i.e., $g_{s,\alpha\beta,i}^{\rm FS}$ is equivalent to $g_{s,\alpha\beta}(\boldsymbol{k_i})$.
Since the real space Hamiltonian $\mathcal{H}^0_{\text{rs}}(\boldsymbol{\phi})$ is unitarily equivalent to the full momentum-space Hamiltonian defined over all discrete momentum points, $\mathcal{H}^{\text{full}}_{\text{ks}}(\boldsymbol{\phi}) = \bigoplus_{\boldsymbol{k_i}} \mathcal{H}^0_{\boldsymbol{k_i+\phi}}$, where $\boldsymbol{k_i}=\left(\frac{2\pi(i_1-1)}{L_1},\cdots,\frac{2\pi(i_d-1)}{L_d}\right),\ i_n=1,\cdots,L_n$. Consequently, $g_{s,\alpha\beta}(\boldsymbol{k_i})$ can be expressed in the same flux-space QM form as Eq.~(\ref{eqgni}) associated with $\mathcal{H}^{\text{full}}_{\text{ks}}(\boldsymbol{\phi})$ by comparing the eigenvectors for $\boldsymbol{\phi}=0$ and $\boldsymbol{\phi}\rightarrow 0$ (see Supplemental Material (SM)\cite{SM}).

Moreover, the integrated flux-space quantum metric (IFSQM) for a selected eigensector $s$ can be directly obtained as $\mathcal{Q}^{\rm FS}_{s,\alpha\beta} = \frac{(2\pi)^{d-1}}{V}\sum_{i}g_{s,\alpha\beta,i}^{\rm FS}$, where $V = \prod_{n=1}^{d} L_n$ denotes the volume. We note that the IFSQM shows excellent agreement with the real-space integrated quantum metric proposed in previous works~\cite{Resta2018,PhysRevLett.122.166602,PhysRevB.107.205133,PhysRevB.111.134201}. Therefore, we establish a flux-space framework to formulate the QM directly in real-space representation, providing a unified approach applicable to both periodic and aperiodic systems. 
In addition, the flux-space QM and IFSQM give almost identical results for open and closed boundary conditions, since the difference between them only constitutes a minor boundary perturbation that vanishes in the thermodynamic limit.


\textit{\textcolor{blue}{Weak-coupling geometric superfluid weight in QCs with isolated flat-energy spectra.--}} 
Using the flux-space QM formulated in real space representation, we investigate the geometric origin of the superfluid weight in QCs and its connection to the flux-space QM, starting from the following attractive Hubbard Hamiltonian,
\begin{align}\label{eqhub1}
\begin{split}
H=H_0-\mu\sum_{i\sigma} c_{i\sigma}^\dagger c_{i\sigma}-U\sum_{i}c_{i\downarrow}^\dagger c_{i\uparrow}^\dagger c_{i\uparrow} c_{i\downarrow},
\end{split}
\end{align}
where $H_0$ denotes the non-interacting hopping Hamiltonian on a general quasicrystal hosting a subset of isolated flat-energy states, $\mu$ is the chemical potential, $\sigma=\uparrow,\downarrow$ denotes the spin and $U$ is the strength of the onsite attractive Hubbard interaction. For the subsequent calculation of the superfluid weight, we impose closed boundary conditions (CBC) on \( H_0 \) by connecting the boundaries of a finite size quasicrystal, forming a ring in 1D (or a torus in 2D) systems. In addition, a flux parameter \( \boldsymbol{\phi} \) is introduced via the Peierls substitution. 

We investigate the $s$-wave superconductivity of the Hamiltonian in Eq.~(\ref{eqhub1}). The interaction is treated by using the Bogoliubov-de Gennes (BdG) mean-field method~\cite{SM,PhysRevB.65.014501(2001),zhu2016bogoliubov,PhysRevB.101.144503(2020)}: \(Uc_{i,\downarrow}^\dagger c_{i,\uparrow}^\dagger c_{i,\uparrow} c_{i,\downarrow}\approx c_{i,\downarrow}^\dagger c_{i,\uparrow}^\dagger\Delta_i+\Delta_ic_{i,\uparrow} c_{i,\downarrow}+U\rho_{i,\uparrow} c^\dagger_{i,\downarrow}c_{i,\downarrow}+U\rho_{i,\downarrow}c^\dagger_{i,\uparrow}c_{i,\uparrow}-U\rho_{i,\uparrow}\rho_{i,\downarrow}-{|\Delta_i|^2}/{U}\). Here, the site-dependent pairing and density order parameters are defined as $\Delta_i = U \langle c_{i,\uparrow} c_{i,\downarrow} \rangle = U \tilde{\Delta}_i$ and $\rho_{i\sigma} = \langle c_{i,\sigma}^\dagger c_{i,\sigma} \rangle$, respectively, which are determined self-consistently by minimizing the ground state energy $E_{\rm GS}(\boldsymbol{\phi})$ of the system. The zero temprature superfluid weight, which is associated with the density of Cooper pairs, is an essential quantity for characterizing the superconducting phase, which is given by~\cite{PhysRev.133.A171(1964),PhysRevB.44.6909(1991),PhysRevB.47.7995(1993),PhysRevB.105.134502(2022),PhysRevB.108.L140503(2023)}: $D_{s,\alpha\beta}=\frac{1}{V}\left. \frac{d^2F}{d\phi_\alpha d\phi_\beta}\right|_{\phi_{\alpha,\beta}=0}$, with $F=E_{\rm GS}(\boldsymbol{\phi})+\mu N$ ($N$ is the total number of particles, which depends on the filling).

We then employ a projection method in real space to demonstrate the geometric contribution of the flat-energy states to the superfluid weight in the quasicrystal~\cite{SM}. In this approach, we focus on the case where the flat-energy states are half-filled. When \( U \) is much smaller than the energy gap separating the flat-energy states from other states, the hybridization between different energy sectors can be neglected, and only the contribution from the flat-energy states to the ground-state energy needs to be retained. Consequently, the projected ground-state energy is given by $E^{\rm proj}_{\rm GS}(\boldsymbol{\phi})=\sum_{<0}\lambda^{\rm f}_i(\boldsymbol{\phi})+\sum_{i}U{|\tilde{\Delta}_i|^2}$,
where $\lambda^{\rm f}_i(\boldsymbol{\phi})$ represent the eigenenergies of the superconductor (SC) quasiparticles originating from the flat-energy states. The summation label $<0$ indicates that the sum is taken only over negative eigenvalues. 
Analytically, we can obtain $\lambda_i^{\rm f}(\boldsymbol{\phi})=\pm U\epsilon^{\rm f}_i(\boldsymbol{\phi})$ with $\epsilon^{\rm f}_i(\boldsymbol{\phi})=\sqrt{p^2+d_i(\boldsymbol{\phi})}$, where the energy difference between the flat state energy $\varepsilon_{\rm f}$ and the Fermi energy $\mu$ is written in terms of $U$ as $pU=\mu-\varepsilon_{\rm f}$. $d_i(\boldsymbol{\phi})$ is the $i$-th eigenvalues of the matrix  $M^{\rm tot}(\boldsymbol{\phi}) =P_{\rm f}^\dagger(-\boldsymbol{\phi})\tilde{\Delta}^*P_{\rm f}(\boldsymbol{\phi})P_{\rm f}^\dagger(\boldsymbol{\phi})\tilde{\Delta}P_{\rm f}(-\boldsymbol{\phi})$, $\tilde{\Delta}={\rm diag}(\tilde{\Delta}_{1}, \tilde{\Delta}_2,\cdots,\tilde{\Delta}_{N_t})$ denotes the diagonal matrix of the pairing order parameters and $P_{\rm f}(\boldsymbol{\phi})$ is a matrix of eigenvectors corresponding to the non-interacting flat-energy states under a flux parameter $\boldsymbol{\phi}$.

The superfluid weight, obtained by using the projection method, can be expressed as 
\begin{align}\label{eqdps}
\begin{split}
D_{ps,\alpha\beta}=\left.-\frac{U}{V}\sum_{i}\frac{1}{2\epsilon_i^{\rm f}}\frac{d^2d_i(\boldsymbol{\phi})}{d\phi_\alpha d\phi_\beta}\right|_{\phi_{\alpha,\beta}=0}+\frac{U}{V}C_{\alpha\beta},
\end{split}
\end{align}
where $C_{\alpha\beta}=\sum_i\frac{d^2|\tilde{\Delta}_i|^2}{d\phi_\alpha d\phi_\beta}|_{\phi_{\alpha,\beta}=0}$. 
According to linear response theory, the total superfluid weight generally consists of a conventional contribution from single-band dispersion derivatives, and a geometric contribution from Bloch eigenvector and pairing-order derivatives~\cite{PhysRevB.106.014518,PhysRevB.95.024515}. From Eq.~(\ref{eqdps}), we see that $D_{ps,\alpha\beta}$  arises as a purely geometric contribution associated with the flux-space variations of the flat-energy eigenstates and the pairing order parameter.

We further decompose the matrix $M^{\text{tot}}$ as $M^{\text{tot}}=|\tilde{\Delta}_0|^2M^{\rm uni}+M^{\rm fluc}$, where $M^{\rm uni}(\boldsymbol{\phi}) =P_{\rm f}^\dagger(-\boldsymbol{\phi})P_{\rm f}(\boldsymbol{\phi})P_{\rm f}^{\dagger}(\boldsymbol{\phi})P_{\rm f}(-\boldsymbol{\phi})$, and $\tilde{\Delta}_0=\frac{1}{N_t}\sum_{i}\tilde{\Delta}_i(\boldsymbol{\phi}=0)$. 
Then the first term in Eq.~(\ref{eqdps}), denoted as $D^{(1)}_{ps,\alpha\beta}$, can be decomposed as $D_{ps,\alpha\beta}^{(1)}\approx D_{ps,\alpha\beta}^{\rm qm}+D_{ps,\alpha\beta}^{\rm fluc}$, where
\begin{align}\label{eqdpsqm}
\begin{split}
D_{ps,\alpha\beta}^{\rm qm}=\frac{U|\tilde{\Delta}_0|^2}{2V}\sum_{i}\frac{4g^{\rm FS}_{{\rm f},\alpha\beta,i}}{\epsilon_i^{\rm f}}, 
\end{split}
\end{align}
and $g^{\rm FS}_{{\rm f},\alpha\beta,i} = \left.\frac{1}{4}\frac{d^2}{d\phi_\alpha d\phi_\beta} d_i^{\rm uni}(\boldsymbol{\phi})\right|_{\phi_{\alpha,\beta}=0}$, with $d_i^{\rm uni}(\boldsymbol{\phi})$ the $i$-th eigenvalue of the quantum distance matrix $M^{\rm rs}_{\rm f}=I-M^{\rm uni}(\boldsymbol{\phi})$. The fluctuation contribution is given by $D_{ps,\alpha\beta}^{\rm fluc}=-\left.\frac{U}{V}\sum_{i}\frac{1}{2\epsilon_i^{\rm f}}\frac{d^2}{d\phi_\alpha d\phi_\beta}d^{\rm fluc}_i\right|_{\phi_{\alpha,\beta}=0}$, where $d_i^{\rm fluc}(\boldsymbol{\phi})$ is the $i$-th eigenvalue of the matrix $M^{\rm fluc}(\boldsymbol{\phi})$~\cite{SM}. 

Notably $g^{\mathrm{FS}}_{{\rm f},\alpha\beta,i}$ is identical to the flux-space QM defined in Eq.~(\ref{eqgni}), while the extra prefactor $\tfrac{1}{4}$ arises from the definition of the quantum distance as the distance between $P_{\rm f}(-\boldsymbol{\phi})$ and $P_{\rm f}(\boldsymbol{\phi})$.
Hence, $D_{ps,\alpha\beta}^{\rm qm}$, explicitly expressed in terms of QM in Eq.~(\ref{eqgni}), can be identified as the QM contribution to the superfluid weight.

\textit{\textcolor{blue}{Tunable geometric superfluid weight in various QCs.--}} To demonstrate the validity of our general theory, we investigate superconductivity in a variety of QCs with flat-energy spectra, encompassing both extrinsic and intrinsic types. Taking the 2D Ammann-Beenker-type quasicrystal (AB-QC) [see Fig~.\ref{fig1}(b)], the 2D Lieb quasicrystal (Lieb-QC)~\cite{SM}, and the 1D stub quasicrystal (Stub-QC) [see Fig~.\ref{fig1}(c)] as representative examples, we demonstrate the geometric origin of the superfluid weight associated with flat-energy states.


Let us first consider the intrinsic AB-QC, which can be viewed as a 2D projection of a four-dimensional periodic Lieb lattice with staggered hopping amplitudes. AB-QC exhibits strong and weak hoppings characterized by the parameter $\delta$, reflecting the projected pattern of the periodic staggered hoppings in the four-dimensional Lieb lattice. To conveniently construct a rectangular geometry suitable for applying CBC, we alternatively generate the Ammann-Beenker tiling using the deflation/inflation rules~\cite{https://doi.org/10.1002/ijch.202300119,PhysRevB.102.115125,PhysRevB.104.144511,SM}. The non-interacting Hamiltonian for a finite-size AB-QC lattice with $N_t$ lattice sites can be expressed in real space as $H_0 = \sum_\sigma \hat{C}_\sigma^\dagger \mathcal{H}^0_{\rm rs} \hat{C}_\sigma$, where $\hat{C}_{\sigma}=(\hat{c}_{1,\sigma},\hat{c}_{2,\sigma},\dots,\hat{c}_{N_t,\sigma})^T$ and $\sigma$ denotes the spin index. The energy spectrum is obtained by directly diagonalizing $\mathcal{H}^0_{\rm rs}$ under CBC. It comprises a flat-energy sector independent of \( \delta \), which can be expanded in terms of compact localized states (CLSs)~\cite{ PhysRevB.34.5208,PhysRevB.95.115135,PhysRevB.99.045107}, together with positive- and negative-energy sectors that vary with \( \delta \)~\cite{SM}.
 \begin{figure}[hbpt]
  \includegraphics[width=8.6cm]{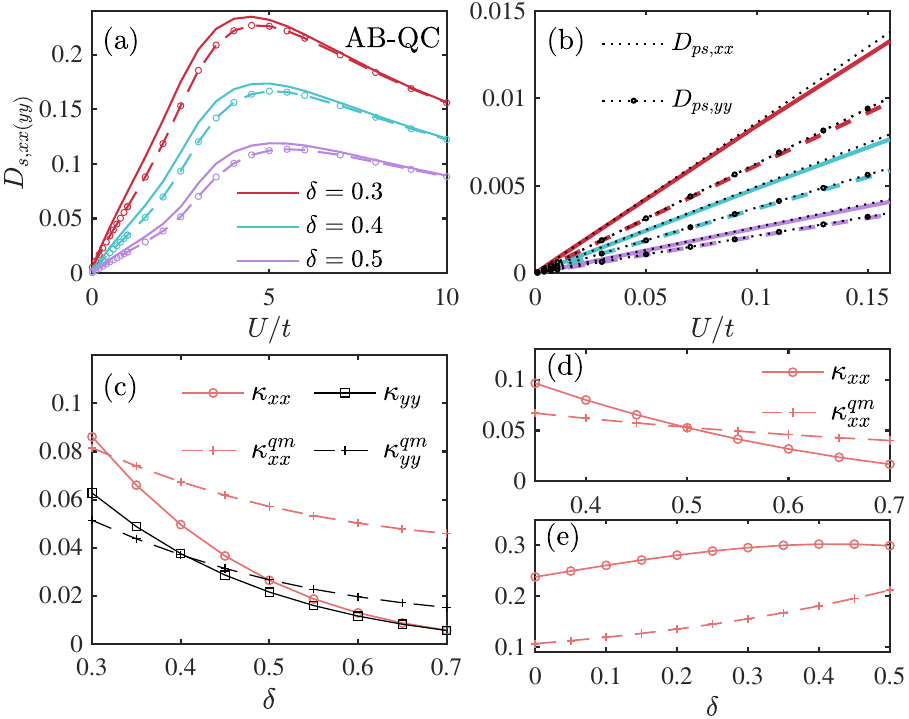}
  \caption{(a) $D_{s,\alpha\alpha}$ of the AB-QC as a function of \( U/t \) for different \( \delta \). Solid (dashed) lines represent the \( xx \) (\( yy \)) component. (b) Comparison of \( D_{ps,\alpha\alpha} \) and \( D_{s,\alpha\alpha} \) in the small-\( U \) regime. Dotted lines corresponds to \( D_{ps,xx} \) and \( D_{ps,yy} \), while the thick solid and dashed lines denote the results in (a). \(\kappa_{\alpha\alpha}\) and \(\kappa_{\alpha\alpha}^{qm}\) for the 2D AB-QC (c), 2D Lieb-QC (d), and 1D Stub-QC (e), plotted as functions of \(\delta\) at \(U=0.01\).
}\label{fig2}
\end{figure}
With the explicit form of $H_0$, the corresponding attractive Hubbard Hamiltonian as in Eq.~(\ref{eqhub1}) is solved by BdG mean-field method~\cite{PhysRevB.65.014501(2001),zhu2016bogoliubov,PhysRevB.101.144503(2020)}. We focus on the ground state at half-filling of the flat-energy spectrum (the filling $\nu \equiv \sum_{i,\sigma} \rho_{i\sigma} / (2N_t) = 1/2$). Due to the quasiperiodic structure, the self-consistently determined pairing amplitude $\Delta_i$ develops a quasiperiodic profile, while $\rho_{i\sigma}$ remains uniform, consistent with the uniform density theorem for bipartite lattices~\cite{Uniformdensitytheorem}. 

In the small-\( U \) regime, the projected result \( D_{ps,\alpha\beta} \) [Eq.~(\ref{eqdps})] accurately reproduces both the $U$-linear behavior and the magnitude of \( D_{s,\alpha\beta} \) [see Fig.~\ref{fig2}(a) and (b)], revealing that the superfluid weight in this regime originates from a purely geometric contribution.
 Owing to the anisotropy between the \( x \) and \( y \) directions, the diagonal components of the superfluid weight, \( D_{s,xx} \) and \( D_{s,yy} \), differ in magnitude, and the off-diagonal component, \( D_{s,xy}=D_{s,yx} \), has very small effect~\cite{SM}. In the small-\( U \) limit where $D_{s,\alpha\alpha}$ is $U$-linear, we extract the slope of the projected superfluid weight \( D_{ps,\alpha\alpha} \), as well as that of the quantum-metric-related component \( D_{ps,\alpha\alpha}^{\rm qm} \) [Eq.~(\ref{eqdpsqm})], denoted respectively as $\kappa_{\alpha\alpha}(\delta)=dD_{s,\alpha\alpha}/dU$ and $\kappa_{\alpha\alpha}^{qm}(\delta)=dD_{ps,\alpha\alpha}^{\rm qm}/dU$. 
 As shown in Fig.~\ref{fig2}(c), both \( \kappa_{\alpha\alpha} \) and \( \kappa_{\alpha\alpha}^{\rm qm} \) exhibit a similar decreasing trend as \( \delta \) increases. The quantum-metric-related term \( D_{ps,\alpha\alpha}^{\rm qm} \) always provides a substantial positive contribution to the superfluid weight, whereas the remaining term, \( [D_{ps,\alpha\alpha} - D_{ps,\alpha\alpha}^{\rm qm}] \), either yields a small positive contribution (when \( \kappa_{\alpha\alpha} > \kappa_{\alpha\alpha}^{qm} \)) or becomes negative, leading to \( D_{ps,\alpha\alpha} < D_{ps,\alpha\alpha}^{\rm qm} \) (when \( \kappa_{\alpha\alpha} < \kappa_{\alpha\alpha}^{\rm qm} \)).

When \( U \) is no longer much smaller than the energy gap between the flat-energy states and the other states, \( D_{s,\alpha\beta} \) includes contributions from the non-flat-energy sectors as well as hybridization between the flat-energy and non-flat-energy sectors.
 In this case, the projected \( D_{ps,\alpha\beta} \) obtained from the isolated flat-energy sector deviates from the total superfluid weight \( D_{s,\alpha\beta} \). In the strong-coupling regime, the superfluid weight decreases with increasing \( U \), following a \( 1/U \) dependence (\( D_{s,\alpha\alpha} \propto 1/U \)). In this limit, the behavior is no longer governed by the quantum metric but is instead associated with the formation of tightly bound pairs in the BEC regime of the BCS-BEC crossover~\cite{PhysRevA.99.053608,PhysRevB.106.014504}. 

We have also constructed models for the Stub-QC and Lieb-QC~\cite{SM} by introducing a quasiperiodic modulation of the hopping amplitudes following the Fibonacci sequence $F_n$~\cite{Fn,RevModPhys.93.045001(2021)}. Similar to the results for the AB-QC, Eq.~(\ref{eqdps}) correctly predicts both the $U$-linear behavior and magnitude of the relevant superfluid weight in the small-\(U\) regime~\cite{SM}. For the Lieb-QC, as shown in Fig.~\ref{fig2}(d), similar to the AB-QC, the quantum-metric-related term \( D_{ps,xx}^{\rm qm} \) always provides a significant positive contribution to the superfluid weight, whereas the remaining term \( [D_{ps,xx} - D_{ps,xx}^{\rm qm}] \) consistently yields either a small positive contribution or a negative one. For the 1D Stub-QC (see Fig.~\ref{fig2}(e)), the increase of \( D_{ps,xx} \) (or \( \kappa_{xx} \)) with \( \delta \) is primarily driven by the corresponding rise in the quantum-metric-related term \( D_{ps,xx}^{\rm qm} \) (or \( \kappa_{xx}^{qm} \)).

\begin{figure}[hbpt]
  \includegraphics[width=8.7cm]{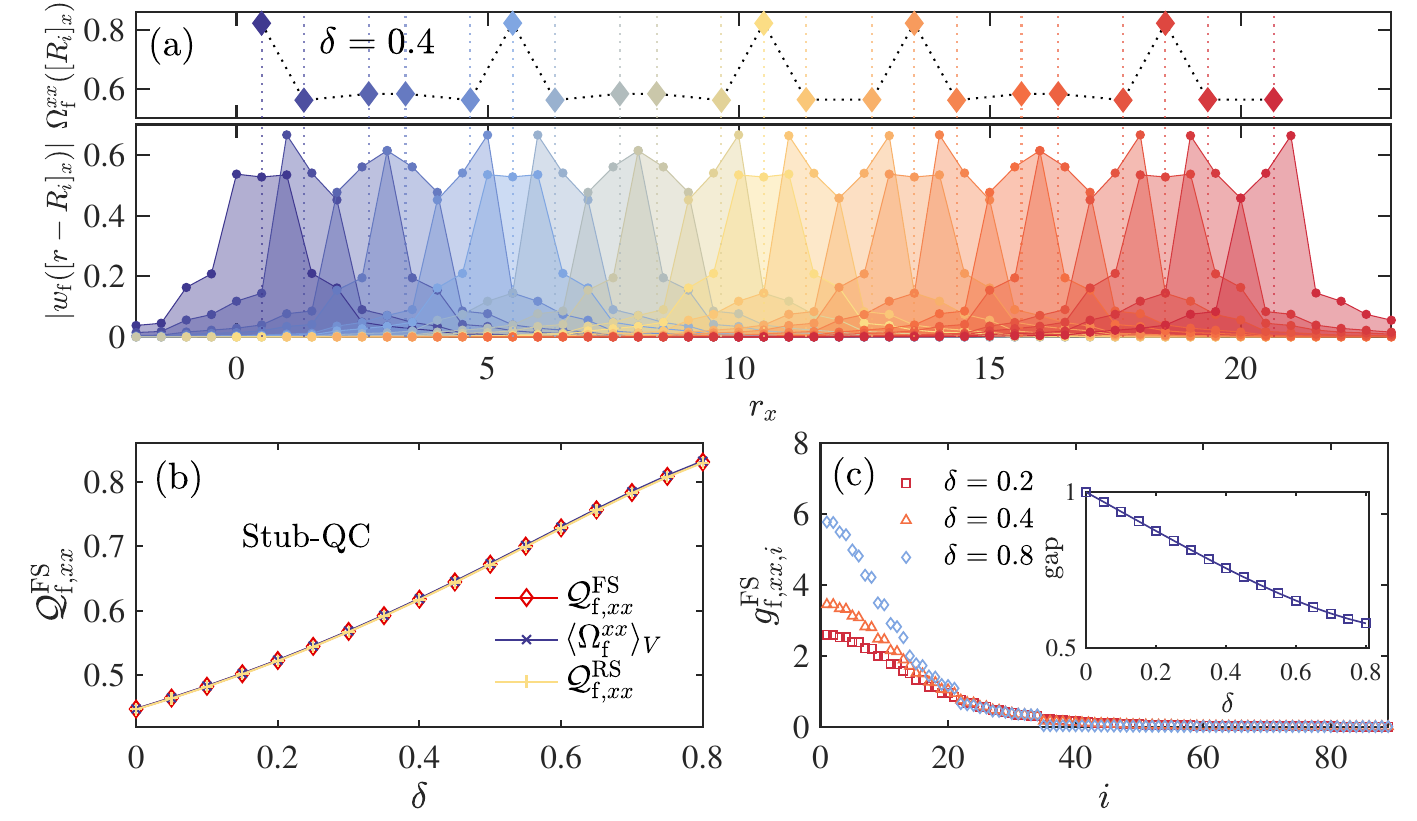}
  \caption{(a) Top panel: Spread function of maximally localized Wannier functions for the 1D Stub-QC at $\delta=0.4$. Bottom panel: Spatial profiles of the Wannier functions for the 1D Stub-QC. Dashed vertical lines indicate the positions of the Wannier centers. The color represents the correspondence between the two panels. (b) $Q^{\rm FS}_{{\rm f},xx}$, \(\langle\Omega^{xx}_{\rm f}\rangle_V\), and \( \mathcal{Q}^{\rm RS}_{{\rm f},xx}\) of the flat-energy sector as functions of \( \delta \) for the 1D Stub-QC. (c) $g^{\rm FS}_{{\rm f},xx,i}$ of the 1D Stub-QC at several different values of $\delta$, with the inset showing the energy gap between the flat-energy sector and the other sectors as a function of $\delta$. 
}\label{fig3}
\end{figure}

\textit{\textcolor{blue}{Quantum metric and Wannier functions.--}} 
Quantum geometry has been shown to effectively characterize the localization properties of Wannier functions. For a general \(d\)-dimensional system, the maximally localized Wannier functions along a given direction \(\alpha\) can be obtained by diagonalizing the projected position operator \(\hat{P} \hat{r}_\alpha \hat{P}\) within a selected eigenstate sector \(s\)~\cite{wfdisordersoliton(1982),SM}, where \(\hat{P}=\sum_{i\in s}|\varphi_i\rangle\langle\varphi_i|\) is the projection operator onto the chosen sector, and \(\hat{r}_\alpha\) is the position operator along the \(\alpha\) direction. The spread function of the Wannier functions along $\alpha$ is then given by $\Omega_{\rm s}^{\alpha\alpha}([R_i]_\alpha)=\langle r_\alpha^2\rangle_s-\langle r_\alpha\rangle_s^2$, where the average $\langle\cdot\rangle_{\rm s}$ is taken over the Wannier function \(w_s([{r} - {R}_i]_\alpha)\), corresponding to the $s$ sector, centered at $[R_i]_\alpha$. We find that the IFSQM is equivalent to the volume-averaged spread function, denoted as $\langle\Omega^{\alpha\alpha}_{\rm f}\rangle_V= 2\frac{(2\pi)^{n-1}}{V}\sum_{R_i}\Omega^{\alpha\alpha}_{\rm f}([R_i]_\alpha)$, and is consistent with the real-space integrated quantum metric (RSIQM) $\mathcal{Q}^{\rm RS}_{s,\alpha\beta}=-\frac{(2\pi)^{n-1}}{V}{\rm Re}\ {\rm Tr}\left\{\hat{P}\left[\hat{r}_\alpha,\hat{P}\right]\left[\hat{r}_\beta,\hat{P}\right]\right\}$ ($\hat{P}$ is the projection operator onto the \( s \)-sector, defined in the same way as before) reported in previous works~\cite{Resta2018,PhysRevLett.122.166602,PhysRevB.107.205133,PhysRevB.111.134201}, 
i.e., \(Q^{\rm FS}_{s,\alpha\alpha} = \langle\Omega^{\alpha\alpha}_{s}\rangle_V=\mathcal{Q}^{\rm RS}_{s,\alpha\alpha}\). Thereby establishing a direct relation between the IFSQM and the Wannier function. 

For clarity, we take the 1D Stub-QC shown in Fig.~\ref{fig1}(c) as a representative example. In this system, the horizontal hopping is uniform, $t_{ij}^{BC} = t$, while the vertical hopping $t_{ij}^{AB}$ is quasiperiodically modulated by the Fibonacci sequence as $t_{ij}^{AB} = [\gamma - F_n(x_B)\delta] t$, where $\delta$ controls the modulation strength, \(x_B\) denotes the integer \(x\)-coordinate of the $B$ site and $F_n(x_B)$ is the $x_B$-th number in the Fibonacci sequence \(F_n\)~\cite{Fn}. Since each $F_n(x_B)$ is either 0 or 1, the vertical hopping $t_{ij}^{AB}$ takes values $\gamma t$ or $(\gamma - \delta)t$. As shown in Fig.~\ref{fig3}(a), the quasiperiodicity breaks the translational symmetry of the Wannier function's spatial profile, producing a quasiperiodic pattern in its spread functions \(\Omega_{s}^{\alpha\alpha}([R_i]_\alpha)\). In this case, the IFSQM captures the volume-averaged properties of the Wannier function spread $\langle\Omega^{\alpha\alpha}_{\rm f}\rangle_V$, and is fully consistent with the results of the RSIQM [see Fig.~\ref{fig3}(b)]. In addition, Fig.~\ref{fig3}(c) shows the flux-space QM of the flat-energy sector at various $\delta$. Its magnitude exhibits an inverse relationship with the energy gap between the flat-energy sector and the other sectors (inset in Fig.~\ref{fig3}(c)), similar to the inverse relationship between the QM and the band gap observed in momentum space. Moreover, the equivalence among the IFSQM, the volume-averaged Wannier function spread, and the RSIQM also holds for the 2D Lieb-QC and AB-QC~\cite{SM}.

\textit{\textcolor{blue}{Conclusion and Discussion.--}}
We have formulated the concept of QM in systems without translational symmetry by introducing a flux-space formulation, thereby providing a foundation for exploring quantum geometric effects in aperiodic systems. In the weak-coupling limit, we deduce the geometric origin of the superfluid weight in QCs with flat-energy spectra. 
We further apply this real-space framework to various extrinsic and intrinsic QCs hosting flat energy spectra, demonstrating the essential role of quantum geometry in flat-energy-spectra superconductivity.
It is straightforward to apply our theory to general QCs in arbitrary dimensions by introducing additional flux variables consistent with spatial dimensions. Revealing the geometric superfluid weight of other intrinsic QCs~\cite{PhysRevB.102.064210(2020),PhysRevB.100.014510,PhysRevLett.134.206001(2025),Diaz-Reynoso_2024,zhang2024FBQC} as well as extrinsic ones through layer twisting~\cite{PhysRevB.99.165430,PhysRevB.104.165112(2021),Zhou2023,Hao2024} is one important direction for future study.

\begin{acknowledgments}
The authors thank Rasoul Ghadimi for
helpful discussions. J.S. acknowledges support from China Scholarship Council and the Academic Excellence Foundation of BUAA for PhD Students. H.G. acknowledge support from the NSFC grant No.~12574249 and the BNLCMP open research fund under Grant No.~2024BNLCMPKF023. B.-J.Y was supported by Samsung Science and Technology Foundation under project no. SSTF-BA2002-06, National Research Foundation of Korea (NRF) funded by the Korean government(MSIT), grant no. RS-2021-NR060087 and RS-2025-00562579, Global Research Development-Center (GRDC) Cooperative Hub Program through the NRF funded by the MSIT, grant no. RS-2023-00258359, Global-LAMP program of the NRF funded by the Ministry of Education, grant no. RS-2023-00301976.
\end{acknowledgments}

\bibliographystyle{apsrev4-1-etal-title_10authors}
\bibliography{ref}

@Article{Torma2022,
author={T{\"o}rm{\"a}, P{\"a}ivi
and Peotta, Sebastiano
and Bernevig, Bogdan A.},
title={Superconductivity, superfluidity and quantum geometry in twisted multilayer systems},
journal={Nature Reviews Physics},
year={2022},
month={Aug},
day={01},
volume={4},
number={8},
pages={528-542},
issn={2522-5820},
doi={10.1038/s42254-022-00466-y},
url={https://doi.org/10.1038/s42254-022-00466-y}
}

@article{PhysRevB.106.014518,
  title = {Revisiting flat band superconductivity: Dependence on minimal quantum metric and band touchings},
  author = {Huhtinen, Kukka-Emilia and Herzog-Arbeitman, Jonah and Chew, Aaron and Bernevig, Bogdan A. and T\"orm\"a, P\"aivi},
  journal = {Phys. Rev. B},
  volume = {106},
  issue = {1},
  pages = {014518},
  numpages = {23},
  year = {2022},
  month = {Jul},
  publisher = {American Physical Society},
  doi = {10.1103/PhysRevB.106.014518},
  url = {https://link.aps.org/doi/10.1103/PhysRevB.106.014518}
}

@article{PhysRevB.95.024515,
  title = {Band geometry, Berry curvature, and superfluid weight},
  author = {Liang, Long and Vanhala, Tuomas I. and Peotta, Sebastiano and Siro, Topi and Harju, Ari and T\"orm\"a, P\"aivi},
  journal = {Phys. Rev. B},
  volume = {95},
  issue = {2},
  pages = {024515},
  numpages = {16},
  year = {2017},
  month = {Jan},
  publisher = {American Physical Society},
  doi = {10.1103/PhysRevB.95.024515},
  url = {https://link.aps.org/doi/10.1103/PhysRevB.95.024515}
}

@article{PhysRevB.109.214518,
  title = {Geometric superfluid weight of composite bands in multiorbital superconductors},
  author = {Jiang, Guodong and Barlas, Yafis},
  journal = {Phys. Rev. B},
  volume = {109},
  issue = {21},
  pages = {214518},
  numpages = {19},
  year = {2024},
  month = {Jun},
  publisher = {American Physical Society},
  doi = {10.1103/PhysRevB.109.214518},
  url = {https://link.aps.org/doi/10.1103/PhysRevB.109.214518}
}

@article{PhysRevA.103.043329,
  title = {Geometry and superfluidity of the flat band in a non-Hermitian optical lattice},
  author = {He, Peng and Ding, Hai-Tao and Zhu, Shi-Liang},
  journal = {Phys. Rev. A},
  volume = {103},
  issue = {4},
  pages = {043329},
  numpages = {10},
  year = {2021},
  month = {Apr},
  publisher = {American Physical Society},
  doi = {10.1103/PhysRevA.103.043329},
  url = {https://link.aps.org/doi/10.1103/PhysRevA.103.043329}
}

@article{PhysRevLett.128.087002,
  title = {Superfluid Weight Bounds from Symmetry and Quantum Geometry in Flat Bands},
  author = {Herzog-Arbeitman, Jonah and Peri, Valerio and Schindler, Frank and Huber, Sebastian D. and Bernevig, B. Andrei},
  journal = {Phys. Rev. Lett.},
  volume = {128},
  issue = {8},
  pages = {087002},
  numpages = {8},
  year = {2022},
  month = {Feb},
  publisher = {American Physical Society},
  doi = {10.1103/PhysRevLett.128.087002},
  url = {https://link.aps.org/doi/10.1103/PhysRevLett.128.087002}
}

@misc{peotta2023,
      title={Quantum geometry in superfluidity and superconductivity}, 
      author={Sebastiano Peotta and Kukka-Emilia Huhtinen and T\"orm\"a, P\"aivi},
      year={2023},
      eprint={2308.08248},
      archivePrefix={arXiv},
      primaryClass={cond-mat.quant-gas},
      url={https://arxiv.org/abs/2308.08248}, 
}

@article{PhysRevLett.123.237002,
  title = {Geometric and Conventional Contribution to the Superfluid Weight in Twisted Bilayer Graphene},
  author = {Hu, Xiang and Hyart, Timo and Pikulin, Dmitry I. and Rossi, Enrico},
  journal = {Phys. Rev. Lett.},
  volume = {123},
  issue = {23},
  pages = {237002},
  numpages = {6},
  year = {2019},
  month = {Dec},
  publisher = {American Physical Society},
  doi = {10.1103/PhysRevLett.123.237002},
  url = {https://link.aps.org/doi/10.1103/PhysRevLett.123.237002}
}

@article{PhysRevLett.124.167002,
  title = {Topology-Bounded Superfluid Weight in Twisted Bilayer Graphene},
  author = {Xie, Fang and Song, Zhida and Lian, Biao and Bernevig, B. Andrei},
  journal = {Phys. Rev. Lett.},
  volume = {124},
  issue = {16},
  pages = {167002},
  numpages = {6},
  year = {2020},
  month = {Apr},
  publisher = {American Physical Society},
  doi = {10.1103/PhysRevLett.124.167002},
  url = {https://link.aps.org/doi/10.1103/PhysRevLett.124.167002}
}

@article{liutianyu2024,
    author = {Liu, Tianyu and Qiang, Xiao-Bin and Lu, Hai-Zhou and Xie, X C},
    title = {Quantum geometry in condensed matter},
    journal = {National Science Review},
    volume = {12},
    number = {3},
    pages = {nwae334},
    year = {2024},
    month = {09},
    issn = {2095-5138},
    doi = {10.1093/nsr/nwae334},
    url = {https://doi.org/10.1093/nsr/nwae334}
}

@article{PhysRevB.94.245149,
  title = {Effective theory and emergent $\text{SU}(2)$ symmetry in the flat bands of attractive Hubbard models},
  author = {Tovmasyan, Murad and Peotta, Sebastiano and T\"orm\"a, P\"aivi and Huber, Sebastian D.},
  journal = {Phys. Rev. B},
  volume = {94},
  issue = {24},
  pages = {245149},
  numpages = {17},
  year = {2016},
  month = {Dec},
  publisher = {American Physical Society},
  doi = {10.1103/PhysRevB.94.245149},
  url = {https://link.aps.org/doi/10.1103/PhysRevB.94.245149}
}

@article{PhysRevB.96.064511,
  title = {Wave-packet dynamics of Bogoliubov quasiparticles: Quantum metric effects},
  author = {Liang, Long and Peotta, Sebastiano and Harju, Ari and T\"orm\"a, P\"aivi},
  journal = {Phys. Rev. B},
  volume = {96},
  issue = {6},
  pages = {064511},
  numpages = {14},
  year = {2017},
  month = {Aug},
  publisher = {American Physical Society},
  doi = {10.1103/PhysRevB.96.064511},
  url = {https://link.aps.org/doi/10.1103/PhysRevB.96.064511}
}

@article{PhysRevB.98.134513,
  title = {Preformed pairs in flat Bloch bands},
  author = {Tovmasyan, Murad and Peotta, Sebastiano and Liang, Long and T\"orm\"a, P\"aivi and Huber, Sebastian D.},
  journal = {Phys. Rev. B},
  volume = {98},
  issue = {13},
  pages = {134513},
  numpages = {22},
  year = {2018},
  month = {Oct},
  publisher = {American Physical Society},
  doi = {10.1103/PhysRevB.98.134513},
  url = {https://link.aps.org/doi/10.1103/PhysRevB.98.134513}
}

@article{PhysRevB.105.024502,
  title = {Pairing and superconductivity in quasi-one-dimensional flat-band systems: Creutz and sawtooth lattices},
  author = {Chan, Si Min and Gr\'emaud, B. and Batrouni, G. G.},
  journal = {Phys. Rev. B},
  volume = {105},
  issue = {2},
  pages = {024502},
  numpages = {10},
  year = {2022},
  month = {Jan},
  publisher = {American Physical Society},
  doi = {10.1103/PhysRevB.105.024502},
  url = {https://link.aps.org/doi/10.1103/PhysRevB.105.024502}
}

@article{PhysRevLett.122.166602,
  title = {Local Theory of the Insulating State},
  author = {Marrazzo, Antimo and Resta, Raffaele},
  journal = {Phys. Rev. Lett.},
  volume = {122},
  issue = {16},
  pages = {166602},
  numpages = {5},
  year = {2019},
  month = {Apr},
  publisher = {American Physical Society},
  doi = {10.1103/PhysRevLett.122.166602},
  url = {https://link.aps.org/doi/10.1103/PhysRevLett.122.166602}
}

@Article{Resta2018,
author={Resta, Raffaele},
title={Theory of the insulating state},
journal={La Rivista del Nuovo Cimento},
year={2018},
month={Sep},
day={01},
volume={41},
number={9},
pages={463-512},
issn={1826-9850},
doi={10.1393/ncr/i2018-10151-1},
url={https://doi.org/10.1393/ncr/i2018-10151-1}
}

@article{PhysRevB.111.134201,
  title = {Scaling of the integrated quantum metric in disordered topological phases},
  author = {Romeral, Jorge Mart\'{\i}nez and Cummings, Aron W. and Roche, Stephan},
  journal = {Phys. Rev. B},
  volume = {111},
  issue = {13},
  pages = {134201},
  numpages = {8},
  year = {2025},
  month = {Apr},
  publisher = {American Physical Society},
  doi = {10.1103/PhysRevB.111.134201},
  url = {https://link.aps.org/doi/10.1103/PhysRevB.111.134201}
}

@article{PhysRevB.107.205133,
  title = {Mapping quantum geometry and quantum phase transitions to real space by a fidelity marker},
  author = {de Sousa, Matheus S. M. and Cruz, Antonio L. and Chen, Wei},
  journal = {Phys. Rev. B},
  volume = {107},
  issue = {20},
  pages = {205133},
  numpages = {11},
  year = {2023},
  month = {May},
  publisher = {American Physical Society},
  doi = {10.1103/PhysRevB.107.205133},
  url = {https://link.aps.org/doi/10.1103/PhysRevB.107.205133}
}

@Article{Resta2011,
author={Resta, R.},
title={The insulating state of matter: a geometrical theory},
journal={The European Physical Journal B},
year={2011},
month={Jan},
day={01},
volume={79},
number={2},
pages={121-137},
issn={1434-6036},
doi={10.1140/epjb/e2010-10874-4},
url={https://doi.org/10.1140/epjb/e2010-10874-4}
}

@article{PhysRevLett.131.240001,
  title = {Essay: Where Can Quantum Geometry Lead Us?},
  author = {T\"orm\"a, P\"aivi},
  journal = {Phys. Rev. Lett.},
  volume = {131},
  issue = {24},
  pages = {240001},
  numpages = {7},
  year = {2023},
  month = {Dec},
  publisher = {American Physical Society},
  doi = {10.1103/PhysRevLett.131.240001},
  url = {https://link.aps.org/doi/10.1103/PhysRevLett.131.240001}
}

@Article{Provost1980,
author={Provost, J. P.
and Vallee, G.},
title={Riemannian structure on manifolds of quantum states},
journal={Communications in Mathematical Physics},
year={1980},
month={Sep},
day={01},
volume={76},
number={3},
pages={289-301},
issn={1432-0916},
doi={10.1007/BF02193559},
url={https://doi.org/10.1007/BF02193559}
}

@article{PhysRevB.94.125408,
  title = {Generalized Aubry-Andr\'e-Harper model with $p$-wave superconducting pairing},
  author = {Zeng, Qi-Bo and Chen, Shu and L\"u, Rong},
  journal = {Phys. Rev. B},
  volume = {94},
  issue = {12},
  pages = {125408},
  numpages = {8},
  year = {2016},
  month = {Sep},
  publisher = {American Physical Society},
  doi = {10.1103/PhysRevB.94.125408},
  url = {https://link.aps.org/doi/10.1103/PhysRevB.94.125408}
}

@article{PhysRevB.93.104504,
  title = {Phase diagram of a non-Abelian Aubry-Andr\'e-Harper model with $p$-wave superfluidity},
  author = {Wang, Jun and Liu, Xia-Ji and Xianlong, Gao and Hu, Hui},
  journal = {Phys. Rev. B},
  volume = {93},
  issue = {10},
  pages = {104504},
  numpages = {7},
  year = {2016},
  month = {Mar},
  publisher = {American Physical Society},
  doi = {10.1103/PhysRevB.93.104504},
  url = {https://link.aps.org/doi/10.1103/PhysRevB.93.104504}
}

@article{PhysRevB.100.014510,
  title = {Conventional superconductivity in quasicrystals},
  author = {Ara\'ujo, Ronaldo N. and Andrade, Eric C.},
  journal = {Phys. Rev. B},
  volume = {100},
  issue = {1},
  pages = {014510},
  numpages = {10},
  year = {2019},
  month = {Jul},
  publisher = {American Physical Society},
  doi = {10.1103/PhysRevB.100.014510},
  url = {https://link.aps.org/doi/10.1103/PhysRevB.100.014510}
}

@article{PhysRevB.102.115108,
  title = {Physical properties of weak-coupling quasiperiodic superconductors},
  author = {Takemori, Nayuta and Arita, Ryotaro and Sakai, Shiro},
  journal = {Phys. Rev. B},
  volume = {102},
  issue = {11},
  pages = {115108},
  numpages = {8},
  year = {2020},
  month = {Sep},
  publisher = {American Physical Society},
  doi = {10.1103/PhysRevB.102.115108},
  url = {https://link.aps.org/doi/10.1103/PhysRevB.102.115108}
}

@article{PhysRevResearch.3.023195,
  title = {Enhanced superconductivity in quasiperiodic crystals},
  author = {Fan, Zhijie and Chern, Gia-Wei and Lin, Shi-Zeng},
  journal = {Phys. Rev. Res.},
  volume = {3},
  issue = {2},
  pages = {023195},
  numpages = {12},
  year = {2021},
  month = {Jun},
  publisher = {American Physical Society},
  doi = {10.1103/PhysRevResearch.3.023195},
  url = {https://link.aps.org/doi/10.1103/PhysRevResearch.3.023195}
}

@article{PhysRevResearch.5.043164,
  title = {Supercurrent distribution in real-space and anomalous paramagnetic response in a superconducting quasicrystal},
  author = {Fukushima, Takumi and Takemori, Nayuta and Sakai, Shiro and Ichioka, Masanori and Jagannathan, Anuradha},
  journal = {Phys. Rev. Res.},
  volume = {5},
  issue = {4},
  pages = {043164},
  numpages = {13},
  year = {2023},
  month = {Nov},
  publisher = {American Physical Society},
  doi = {10.1103/PhysRevResearch.5.043164},
  url = {https://link.aps.org/doi/10.1103/PhysRevResearch.5.043164}
}

@article{PhysRevLett.133.136002,
  title = {Nematic Superconductivity and Its Critical Vestigial Phases in the Quasicrystal},
  author = {Liu, Yu-Bo and Zhou, Jing and Yang, Fan},
  journal = {Phys. Rev. Lett.},
  volume = {133},
  issue = {13},
  pages = {136002},
  numpages = {8},
  year = {2024},
  month = {Sep},
  publisher = {American Physical Society},
  doi = {10.1103/PhysRevLett.133.136002},
  url = {https://link.aps.org/doi/10.1103/PhysRevLett.133.136002}
}

@article{PhysRevB.110.134508,
  title = {Topological superconductivity in Fibonacci quasicrystals},
  author = {Kobia\l{}ka, Aksel and Awoga, Oladunjoye A. and Leijnse, Martin and Doma\ifmmode \acute{n}\else \'{n}\fi{}ski, Tadeusz and Holmvall, Patric and Black-Schaffer, Annica M.},
  journal = {Phys. Rev. B},
  volume = {110},
  issue = {13},
  pages = {134508},
  numpages = {20},
  year = {2024},
  month = {Oct},
  publisher = {American Physical Society},
  doi = {10.1103/PhysRevB.110.134508},
  url = {https://link.aps.org/doi/10.1103/PhysRevB.110.134508}
}

@article{PhysRevB.109.134504,
  title = {Enhancement of superconductivity in the Fibonacci chain},
  author = {Sun, Meng and \ifmmode \check{C}\else \v{C}\fi{}ade\ifmmode \check{z}\else \v{z}\fi{}, Tilen and Yurkevich, Igor and Andreanov, Alexei},
  journal = {Phys. Rev. B},
  volume = {109},
  issue = {13},
  pages = {134504},
  numpages = {7},
  year = {2024},
  month = {Apr},
  publisher = {American Physical Society},
  doi = {10.1103/PhysRevB.109.134504},
  url = {https://link.aps.org/doi/10.1103/PhysRevB.109.134504}
}

@Article{Ghadimi2025,
author={Ghadimi, Rasoul
and Yang, Bohm-Jung},
title={Quasiperiodic Pairing in Graphene Quasicrystals},
journal={Nano Letters},
year={2025},
month={Feb},
day={05},
publisher={American Chemical Society},
volume={25},
number={5},
pages={1808-1815},
issn={1530-6984},
doi={10.1021/acs.nanolett.4c04386},
url={https://doi.org/10.1021/acs.nanolett.4c04386}
}

@article{PhysRevB.34.5208,
  title = {Localization of electronic wave functions due to local topology},
  author = {Sutherland, Bill},
  journal = {Phys. Rev. B},
  volume = {34},
  issue = {8},
  pages = {5208--5211},
  numpages = {0},
  year = {1986},
  month = {Oct},
  publisher = {American Physical Society},
  doi = {10.1103/PhysRevB.34.5208},
  url = {https://link.aps.org/doi/10.1103/PhysRevB.34.5208}
}

@article{PhysRevB.95.115135,
  title = {Compact localized states and flat-band generators in one dimension},
  author = {Maimaiti, Wulayimu and Andreanov, Alexei and Park, Hee Chul and Gendelman, Oleg and Flach, Sergej},
  journal = {Phys. Rev. B},
  volume = {95},
  issue = {11},
  pages = {115135},
  numpages = {8},
  year = {2017},
  month = {Mar},
  publisher = {American Physical Society},
  doi = {10.1103/PhysRevB.95.115135},
  url = {https://link.aps.org/doi/10.1103/PhysRevB.95.115135}
}

@article{PhysRevB.99.045107,
  title = {Classification of flat bands according to the band-crossing singularity of Bloch wave functions},
  author = {Rhim, Jun-Won and Yang, Bohm-Jung},
  journal = {Phys. Rev. B},
  volume = {99},
  issue = {4},
  pages = {045107},
  numpages = {21},
  year = {2019},
  month = {Jan},
  publisher = {American Physical Society},
  doi = {10.1103/PhysRevB.99.045107},
  url = {https://link.aps.org/doi/10.1103/PhysRevB.99.045107}
}

@article{zhu2016bogoliubov,
  title={Bogoliubov-de Gennes method and its applications},
  author={Zhu, Jian-Xin},
  volume={924},
  year={2016},
  publisher={Springer}
}

@article{Resta1998PRL,
  title = {Quantum-Mechanical Position Operator in Extended Systems},
  author = {Resta, Raffaele},
  journal = {Phys. Rev. Lett.},
  volume = {80},
  issue = {9},
  pages = {1800--1803},
  numpages = {0},
  year = {1998},
  month = {Mar},
  publisher = {American Physical Society},
  doi = {10.1103/PhysRevLett.80.1800},
  url = {https://link.aps.org/doi/10.1103/PhysRevLett.80.1800}
}

@misc{zhang2024FBQC,
      title={Site-selective correlations in interacting "flat-band" quasicrystals}, 
      author={Yuxi Zhang and Richard T. Scalettar and Rafael M. Fernandes},
      year={2024},
      eprint={2410.07894},
      archivePrefix={arXiv},
      primaryClass={cond-mat.str-el},
      url={https://arxiv.org/abs/2410.07894}, 
}

@Article{Kamiya2018,
author={Kamiya, K.
and Takeuchi, T.
and Kabeya, N.
and Wada, N.
and Ishimasa, T.
and Ochiai, A.
and Deguchi, K.
and Imura, K.
and Sato, N. K.},
title={Discovery of superconductivity in quasicrystal},
journal={Nature Communications},
year={2018},
month={Jan},
day={11},
volume={9},
number={1},
pages={154},
issn={2041-1723},
doi={10.1038/s41467-017-02667-x},
url={https://doi.org/10.1038/s41467-017-02667-x}
}

@Article{Uri2023,
author={Uri, Aviram
and de la Barrera, Sergio C.
and Randeria, Mallika T.
and Rodan-Legrain, Daniel
and Devakul, Trithep
and Crowley, Philip J. D.
and Paul, Nisarga
and Watanabe, Kenji
and Taniguchi, Takashi
and Lifshitz, Ron
and Fu, Liang
and Ashoori, Raymond C.
and Jarillo-Herrero, Pablo},
title={Superconductivity and strong interactions in a tunable moir{\'e} quasicrystal},
journal={Nature},
year={2023},
month={Aug},
day={01},
volume={620},
number={7975},
pages={762-767},
issn={1476-4687},
doi={10.1038/s41586-023-06294-z},
url={https://doi.org/10.1038/s41586-023-06294-z}
}

@Article{Tokumoto2024,
author={Tokumoto, Yuki
and Hamano, Kotaro
and Nakagawa, Sunao
and Kamimura, Yasushi
and Suzuki, Shintaro
and Tamura, Ryuji
and Edagawa, Keiichi},
title={Superconductivity in a van der Waals layered quasicrystal},
journal={Nature Communications},
year={2024},
month={Mar},
day={01},
volume={15},
number={1},
pages={1529},
issn={2041-1723},
doi={10.1038/s41467-024-45952-2},
url={https://doi.org/10.1038/s41467-024-45952-2}
}

@article{Diaz-Reynoso_2024,
doi = {10.1088/1361-648X/ad5acf},
url = {https://dx.doi.org/10.1088/1361-648X/ad5acf},
year = {2024},
month = {jul},
publisher = {IOP Publishing},
volume = {36},
number = {39},
pages = {395502},
author = {Díaz-Reynoso, U A and Huipe-Domratcheva, E and Navarro, O},
title = {Flat-bands in translated and twisted bilayer Penrose quasicrystals},
journal = {Journal of Physics: Condensed Matter}
}

@Misc{SM,
  note = {See Supplemental Material for more details}
}

@misc{yu2025quantumgeometryquantummaterials,
      title={Quantum Geometry in Quantum Materials}, 
      author={Jiabin Yu and B. Andrei Bernevig and Raquel Queiroz and Enrico Rossi and P\“aivi T\"orm\"a and Bohm-Jung Yang},
      year={2025},
      eprint={2501.00098},
      archivePrefix={arXiv},
      primaryClass={cond-mat.mes-hall},
      url={https://arxiv.org/abs/2501.00098}, 
}

@article{RevModPhys.82.1959,
  title = {Berry phase effects on electronic properties},
  author = {Xiao, Di and Chang, Ming-Che and Niu, Qian},
  journal = {Rev. Mod. Phys.},
  volume = {82},
  issue = {3},
  pages = {1959--2007},
  numpages = {0},
  year = {2010},
  month = {Jul},
  publisher = {American Physical Society},
  doi = {10.1103/RevModPhys.82.1959},
  url = {https://link.aps.org/doi/10.1103/RevModPhys.82.1959}
}

@article{RevModPhys.90.015001,
  title = {Weyl and Dirac semimetals in three-dimensional solids},
  author = {Armitage, N. P. and Mele, E. J. and Vishwanath, Ashvin},
  journal = {Rev. Mod. Phys.},
  volume = {90},
  issue = {1},
  pages = {015001},
  numpages = {57},
  year = {2018},
  month = {Jan},
  publisher = {American Physical Society},
  doi = {10.1103/RevModPhys.90.015001},
  url = {https://link.aps.org/doi/10.1103/RevModPhys.90.015001}
}

@article{RevModPhys.82.3045,
  title = {Colloquium: Topological insulators},
  author = {Hasan, M. Z. and Kane, C. L.},
  journal = {Rev. Mod. Phys.},
  volume = {82},
  issue = {4},
  pages = {3045--3067},
  numpages = {0},
  year = {2010},
  month = {Nov},
  publisher = {American Physical Society},
  doi = {10.1103/RevModPhys.82.3045},
  url = {https://link.aps.org/doi/10.1103/RevModPhys.82.3045}
}

@article{RevModPhys.83.1057,
  title = {Topological insulators and superconductors},
  author = {Qi, Xiao-Liang and Zhang, Shou-Cheng},
  journal = {Rev. Mod. Phys.},
  volume = {83},
  issue = {4},
  pages = {1057--1110},
  numpages = {0},
  year = {2011},
  month = {Oct},
  publisher = {American Physical Society},
  doi = {10.1103/RevModPhys.83.1057},
  url = {https://link.aps.org/doi/10.1103/RevModPhys.83.1057}
}

@article{gao2023quantum,
author = {Anyuan Gao  and Yu-Fei Liu  and Jian-Xiang Qiu  and Barun Ghosh  and Thaís V. Trevisan  and Yugo Onishi  and Chaowei Hu  and Tiema Qian  and Hung-Ju Tien  and Shao-Wen Chen  and Mengqi Huang  and Damien Bérubé  and Houchen Li  and Christian Tzschaschel  and Thao Dinh  and Zhe Sun  and Sheng-Chin Ho  and Shang-Wei Lien  and Bahadur Singh  and Kenji Watanabe  and Takashi Taniguchi  and David C. Bell  and Hsin Lin  and Tay-Rong Chang  and Chunhui Rita Du  and Arun Bansil  and Liang Fu  and Ni Ni  and Peter P. Orth  and Qiong Ma  and Su-Yang Xu },
title = {Quantum metric nonlinear Hall effect in a topological antiferromagnetic heterostructure},
journal = {Science},
volume = {381},
number = {6654},
pages = {181-186},
year = {2023},
doi = {10.1126/science.adf1506},
URL = {https://www.science.org/doi/abs/10.1126/science.adf1506},
}

@Article{wang2023quantum,
author={Wang, Naizhou
and Kaplan, Daniel
and Zhang, Zhaowei
and Holder, Tobias
and Cao, Ning
and Wang, Aifeng
and Zhou, Xiaoyuan
and Zhou, Feifei
and Jiang, Zhengzhi
and Zhang, Chusheng
and Ru, Shihao
and Cai, Hongbing
and Watanabe, Kenji
and Taniguchi, Takashi
and Yan, Binghai
and Gao, Weibo},
title={Quantum-metric-induced nonlinear transport in a topological antiferromagnet},
journal={Nature},
year={2023},
month={Sep},
day={01},
volume={621},
number={7979},
pages={487-492},
issn={1476-4687},
doi={10.1038/s41586-023-06363-3},
url={https://doi.org/10.1038/s41586-023-06363-3}
}

@Article{tian2023evidence,
author={Tian, Haidong
and Gao, Xueshi
and Zhang, Yuxin
and Che, Shi
and Xu, Tianyi
and Cheung, Patrick
and Watanabe, Kenji
and Taniguchi, Takashi
and Randeria, Mohit
and Zhang, Fan
and Lau, Chun Ning
and Bockrath, Marc W.},
title={Evidence for Dirac flat band superconductivity enabled by quantum geometry},
journal={Nature},
year={2023},
month={Feb},
day={01},
volume={614},
number={7948},
pages={440-444},
issn={1476-4687},
doi={10.1038/s41586-022-05576-2},
url={https://doi.org/10.1038/s41586-022-05576-2}
}

@Article{rhim2020quantum,
author={Rhim, Jun-Won
and Kim, Kyoo
and Yang, Bohm-Jung},
title={Quantum distance and anomalous Landau levels of flat bands},
journal={Nature},
year={2020},
month={Aug},
day={01},
volume={584},
number={7819},
pages={59-63},
issn={1476-4687},
doi={10.1038/s41586-020-2540-1},
url={https://doi.org/10.1038/s41586-020-2540-1}
}

@Article{hwang2021geometric,
author={Hwang, Yoonseok
and Rhim, Jun-Won
and Yang, Bohm-Jung},
title={Geometric characterization of anomalous Landau levels of isolated flat bands},
journal={Nature Communications},
year={2021},
month={Nov},
day={05},
volume={12},
number={1},
pages={6433},
issn={2041-1723},
doi={10.1038/s41467-021-26765-z},
url={https://doi.org/10.1038/s41467-021-26765-z}
}

@article{PhysRevB.109.035134,
  title = {Quantum geometry and Landau levels of quadratic band crossings},
  author = {Jung, Junseo and Lim, Hyeongmuk and Yang, Bohm-Jung},
  journal = {Phys. Rev. B},
  volume = {109},
  issue = {3},
  pages = {035134},
  numpages = {13},
  year = {2024},
  month = {Jan},
  publisher = {American Physical Society},
  doi = {10.1103/PhysRevB.109.035134},
  url = {https://link.aps.org/doi/10.1103/PhysRevB.109.035134}
}

@Article{kang2025measurements,
author={Kang, Mingu
and Kim, Sunje
and Qian, Yuting
and Neves, Paul M.
and Ye, Linda
and Jung, Junseo
and Puntel, Denny
and Mazzola, Federico
and Fang, Shiang
and Jozwiak, Chris
and Bostwick, Aaron
and Rotenberg, Eli
and Fuji, Jun
and Vobornik, Ivana
and Park, Jae-Hoon
and Checkelsky, Joseph G.
and Yang, Bohm-Jung
and Comin, Riccardo},
title={Measurements of the quantum geometric tensor in solids},
journal={Nature Physics},
year={2025},
month={Jan},
day={01},
volume={21},
number={1},
pages={110-117},
issn={1745-2481},
doi={10.1038/s41567-024-02678-8},
url={https://doi.org/10.1038/s41567-024-02678-8}
}

@article{doi:10.1126/science.ado6049,
author = {Sunje Kim  and Yoonah Chung  and Yuting Qian  and Soobin Park  and Chris Jozwiak  and Eli Rotenberg  and Aaron Bostwick  and Keun Su Kim  and Bohm-Jung Yang },
title = {Direct measurement of the quantum metric tensor in solids},
journal = {Science},
volume = {388},
number = {6751},
pages = {1050-1054},
year = {2025},
doi = {10.1126/science.ado6049},
URL = {https://www.science.org/doi/abs/10.1126/science.ado6049},

}

@article{Uniformdensitytheorem,
    author = {Lieb, Elliott H. and Loss, Michael and McCann, Robert J.},
    title = {Uniform density theorem for the Hubbard model},
    journal = {Journal of Mathematical Physics},
    volume = {34},
    number = {3},
    pages = {891-898},
    year = {1993},
    month = {03},
    issn = {0022-2488},
    doi = {10.1063/1.530199},
    url = {https://doi.org/10.1063/1.530199},
    eprint = {https://pubs.aip.org/aip/jmp/article-pdf/34/3/891/19299763/891\_1\_online.pdf},
}

@Article{Hao2024,
author={Hao, Chen-Yue
and Zhan, Zhen
and Pantale{\'o}n, Pierre A.
and He, Jia-Qi
and Zhao, Ya-Xin
and Watanabe, Kenji
and Taniguchi, Takashi
and Guinea, Francisco
and He, Lin},
title={Robust flat bands in twisted trilayer graphene moir{\'e} quasicrystals},
journal={Nature Communications},
year={2024},
month={Sep},
day={30},
volume={15},
number={1},
pages={8437},
issn={2041-1723},
doi={10.1038/s41467-024-52784-7},
url={https://doi.org/10.1038/s41467-024-52784-7}
}

@article{PhysRevB.99.165430,
  title = {Quasicrystalline electronic states in ${30}^{\ensuremath{\circ}}$ rotated twisted bilayer graphene},
  author = {Moon, Pilkyung and Koshino, Mikito and Son, Young-Woo},
  journal = {Phys. Rev. B},
  volume = {99},
  issue = {16},
  pages = {165430},
  numpages = {11},
  year = {2019},
  month = {Apr},
  publisher = {American Physical Society},
  doi = {10.1103/PhysRevB.99.165430},
  url = {https://link.aps.org/doi/10.1103/PhysRevB.99.165430}
}

@Article{Zhou2023,
author={Zhou, Benjamin T.
and Egan, Shannon
and Kush, Dhruv
and Franz, Marcel},
title={Non-Abelian topological superconductivity in maximally twisted double-layer spin-triplet valley-singlet superconductors},
journal={Communications Physics},
year={2023},
month={Mar},
day={15},
volume={6},
number={1},
pages={47},
issn={2399-3650},
doi={10.1038/s42005-023-01165-5},
url={https://doi.org/10.1038/s42005-023-01165-5}
}

@misc{marsal2025QM_quasicrystal,
      title={Quantum metric and localization in a quasicrystal}, 
      author={Quentin Marsal and Patric Holmvall and Annica M. Black-Schaffer},
      year={2025},
      eprint={2506.15575},
      archivePrefix={arXiv},
      primaryClass={cond-mat.mes-hall},
      url={https://arxiv.org/abs/2506.15575}, 
}

@Misc{Fn,
  note = {The Fibonacci sequence is constructed recursively through a symbolic concatenation rule \(F_n = F_{n-2} \bigoplus F_{n-1}\), with the initial conditions \(F_0 = 0\) and \(F_1 = 1\). This generates sequences such as $F_2 = 01, F_3 = 101, F_4 = 01101, \cdots$.}
}

@article{PhysRevB.104.144511,
  title = {Topological superconductivity in quasicrystals},
  author = {Ghadimi, Rasoul and Sugimoto, Takanori and Tanaka, K. and Tohyama, Takami},
  journal = {Phys. Rev. B},
  volume = {104},
  issue = {14},
  pages = {144511},
  numpages = {10},
  year = {2021},
  month = {Oct},
  publisher = {American Physical Society},
  doi = {10.1103/PhysRevB.104.144511},
  url = {https://link.aps.org/doi/10.1103/PhysRevB.104.144511}
}

@article{PhysRevB.102.115125,
  title = {Superlattice structure in the antiferromagnetically ordered state in the Hubbard model on the Ammann-Beenker tiling},
  author = {Koga, Akihisa},
  journal = {Phys. Rev. B},
  volume = {102},
  issue = {11},
  pages = {115125},
  numpages = {10},
  year = {2020},
  month = {Sep},
  publisher = {American Physical Society},
  doi = {10.1103/PhysRevB.102.115125},
  url = {https://link.aps.org/doi/10.1103/PhysRevB.102.115125}
}

@article{https://doi.org/10.1002/ijch.202300119,
author = {Jagannathan, Anuradha and Duneau, Michel},
title = {Properties of the Ammann-Beenker Tiling and its Square Periodic Approximants},
journal = {Israel Journal of Chemistry},
volume = {64},
number = {10-11},
pages = {e202300119},
doi = {https://doi.org/10.1002/ijch.202300119},
url = {https://onlinelibrary.wiley.com/doi/abs/10.1002/ijch.202300119},
year = {2024}
}

@misc{jeon2022discovery,
      title={Discovery of new quasicrystals from translation of hypercubic lattice}, 
      author={Junmo Jeon and SungBin Lee},
      year={2022},
      eprint={2112.14783},
      archivePrefix={arXiv},
      primaryClass={cond-mat.other},
      url={https://arxiv.org/abs/2112.14783}, 
}

@misc{li2025unconventionalalter,
      title={Unconventional Altermagnetism in Quasicrystals: A Hyperspatial Projective Construction}, 
      author={Yiming Li and Mingxiang Pan and Jun Leng and Yuxiao Chen and Huaqing Huang},
      year={2025},
      eprint={2508.01564},
      archivePrefix={arXiv},
      primaryClass={cond-mat.mes-hall},
      url={https://arxiv.org/abs/2508.01564}, 
}

@article{PhysRevA.99.053608,
  title = {Origin of flat-band superfluidity on the Mielke checkerboard lattice},
  author = {Iskin, M.},
  journal = {Phys. Rev. A},
  volume = {99},
  issue = {5},
  pages = {053608},
  numpages = {9},
  year = {2019},
  month = {May},
  publisher = {American Physical Society},
  doi = {10.1103/PhysRevA.99.053608},
  url = {https://link.aps.org/doi/10.1103/PhysRevA.99.053608}
}

@article{PhysRevB.106.014504,
  title = {Pairs, trimers, and BCS-BEC crossover near a flat band: Sawtooth lattice},
  author = {Orso, Giuliano and Singh, Manpreet},
  journal = {Phys. Rev. B},
  volume = {106},
  issue = {1},
  pages = {014504},
  numpages = {15},
  year = {2022},
  month = {Jul},
  publisher = {American Physical Society},
  doi = {10.1103/PhysRevB.106.014504},
  url = {https://link.aps.org/doi/10.1103/PhysRevB.106.014504}
}

@misc{DebikaDebnath2025,
      title={Superconducting order parameter manifested in quasicrystals}, 
      author={Sougata Biswas and Debika Debnath and Paramita Dutta},
      year={2025},
      eprint={2507.14671},
      archivePrefix={arXiv},
      primaryClass={cond-mat.supr-con},
      url={https://arxiv.org/abs/2507.14671}, 
}

@article{MannaSourav2024,
  title = {Noncrystalline topological superconductors},
  author = {Manna, Sourav and Das, Sanjib Kumar and Roy, Bitan},
  journal = {Phys. Rev. B},
  volume = {109},
  issue = {17},
  pages = {174512},
  numpages = {8},
  year = {2024},
  month = {May},
  publisher = {American Physical Society},
  doi = {10.1103/PhysRevB.109.174512},
  url = {https://link.aps.org/doi/10.1103/PhysRevB.109.174512}
}

@article{Balazs2023,
  title = {Fluctuations, uncertainty relations, and the geometry of quantum state manifolds},
  author = {Het\'enyi, Bal\'azs and L\'evay, P\'eter},
  journal = {Phys. Rev. A},
  volume = {108},
  issue = {3},
  pages = {032218},
  numpages = {13},
  year = {2023},
  month = {Sep},
  publisher = {American Physical Society},
  doi = {10.1103/PhysRevA.108.032218},
  url = {https://link.aps.org/doi/10.1103/PhysRevA.108.032218}
}




\end{document}


\title{Supplemental Material for "Geometric Superfluid Weight in Quasicrystals"}
\author{Junsong Sun}
\affiliation{Department of Physics and Astronomy, Seoul National University, Seoul 08826, Korea}
\affiliation{School of Physics, Beihang University,
Beijing, 100191, China}

\author{Huaiming Guo}
\email{hmguo@buaa.edu.cn}
\affiliation{School of Physics, Beihang University,
Beijing, 100191, China}

\author{Bohm-Jung Yang}
\email{bjyang@snu.ac.kr}
\affiliation{Department of Physics and Astronomy, Seoul National University, Seoul 08826, Korea}
\affiliation{Center for Theoretical Physics (CTP), Seoul National University, Seoul 08826, Korea}
\affiliation{Institute of Applied Physics, Seoul National University, Seoul 08826, Korea}
\date{\today}
\begin{abstract}

\end{abstract}
\maketitle
\renewcommand{\thefigure}{S\arabic{figure}}
\renewcommand{\theequation}{S\arabic{equation}}
\renewcommand{\thesection}{S\arabic{section}}
\setcounter{figure}{0}  
\setcounter{equation}{0} 
\numberwithin{equation}{section}\tableofcontents{}

\section{Further discussion on the flux-space quantum metric}
In the main text, when a flux parameter $\boldsymbol{\phi}$ is introduced into real-space Hamiltonian $\mathcal{H}^0_{\text{rs}}(\boldsymbol{\phi})$, we define the quantum distance matrix $M_s^{\rm rs}(\boldsymbol{\phi})$, which characterizes the distance associated with the variation of a selected eigenvector subspace(denote the subspace sector as $s$) $P_s(\boldsymbol{\phi})$ with respect to $\boldsymbol{\phi}$, as follows
\begin{align}\label{eqMrs}
M^{\mathrm{rs}}_s(\boldsymbol{\phi}) = I - P^\dagger_s(0) P_s(\boldsymbol{\phi}) P^\dagger_s(\boldsymbol{\phi}) P_s(0),
\end{align}
where $I$ is an identity matrix, $P_s(\boldsymbol{\phi})=(v_1,v_2,\cdots,v_{N_s})$ is the matrix composed
of the eigenvectors corresponding to the $s$ eigenstate sector of the non-interacting energy spectrum under magnetic flux $\boldsymbol{\phi}$,
$v_i$ denotes the eigenvector corresponding to the $i$-th eigenstate in the $s$ sector, and $N_s$ is the number of eigenstates in the $s$ sector. When $P_s(0)$ and $P_s(\boldsymbol{\phi})$ coincide exactly, the distance matrix reduces to the zero matrix, indicating zero distance. In contrast, when $P_s(0)$ and $P_s(\boldsymbol{\phi})$ are completely orthogonal, $M^{\mathrm{rs}}_s(\boldsymbol{\phi})$ becomes the identity matrix, representing the maximal possible distance between them. Accordingly, our flux-space quantum metric (QM) $g_{s,\alpha\beta,i}^{\rm FS}$ is defined as the metric characterizing the variation of the eigenvalues of the quantum distance matrix $M^{\mathrm{rs}}_s(\boldsymbol{\phi})$
 \begin{align}\label{eqgni}
g_{s,\alpha\beta,i}^{\rm FS}=\left.\frac{d^2}{d\phi_\alpha\phi_\beta}d^{\rm rs}_{s,i}(\boldsymbol{\phi})\right|_{\phi_{\alpha,\beta}=0},
\end{align}
\( d^{\rm rs}_{s,i}(\boldsymbol{\phi}) \) denotes the \( i \)-th eigenvalue of the matrix $M^{\rm rs}_s(\boldsymbol{\phi})$.

Next, we demonstrate that in the finite-size periodic case, the quantum metric in momentum space can be interpreted in the same sense as the flux-space QM defined above, and can be expressed in the same form as Eq.~(\ref{eqgni}). In periodic crystals, the momentum space QM (Fubini-Study metric) characterizes the quantum distance between two infinitesimally close quantum states
($D^2=1-|\langle u_{n\boldsymbol{k}}|u_{s\boldsymbol{k}+d\boldsymbol{k}} \rangle|^2\approx \sum_{i,j}\frac{1}{2}g_{s,\alpha\beta}(\boldsymbol{k})dk_\alpha dk_\beta$)~\cite{Provost1980,liutianyu2024}. 
 Therefore, the momentum space QM can be written as
\begin{align}\label{eqgkBure}
\begin{split}
g_{s,\alpha\beta}(\boldsymbol{k})&\approx \left.\frac{d^2}{d\phi_\alpha d\phi_\beta}[1-|\langle u_s(\boldsymbol{k})|u_s(\boldsymbol{k}+\boldsymbol{\phi})\rangle|^2]\right|_{\phi_{\alpha,\beta}=0}\\ 
&= \left.\frac{d^2}{d\phi_\alpha d\phi_\beta}[1-\langle u_s(\boldsymbol{k})|u_s(\boldsymbol{k}+\boldsymbol{\phi})\rangle\langle u_s(\boldsymbol{k}+\boldsymbol{\phi})|u_s(\boldsymbol{k})\rangle]\right|_{\phi_{\alpha,\beta}=0}\\
&= 2{\rm Re}\{\langle \partial_{k_\alpha} u_{s}(\boldsymbol{k})|[1-|u_{s}(\boldsymbol{k})\rangle\langle u_{s}(\boldsymbol{k})]|\partial_{k_\beta} u_{s}(\boldsymbol{k})\rangle\},
\end{split}
\end{align}
where, $\boldsymbol{\phi}$ serves as an auxiliary parameter representing a small momentum difference. However, it can also be interpreted as a parameter corresponding to magnetic flux, in which case the associated quantum distance measures the change in the state $|u_s(\boldsymbol{k})\rangle$ under a finite flux $\boldsymbol{\phi}$. For simplicity, we take the one-dimensional(1D) case as an example; higher-dimensional cases can be treated in a similar manner.
In a finite-size 1D periodic system of size $L$, the momentum is restricted to a discrete set of points $[k_i=k^x_i = \frac{2\pi(i-1)}{L}, \, i = 1, \cdots, L]$. Accordingly, the QM of the finite-size periodic system is defined at these discrete momenta as $g_{s,xx}(k_i)$. The full momentum-space Hamiltonian at all momentum points is given by
\begin{align}\label{}
\mathcal{H}^{\text{full}}_{\text{ks}}(\boldsymbol{\phi}) = \bigoplus_{i=1}^L \mathcal{H}^0_{k_i+\phi_x},
\end{align}
where $\mathcal{H}^0_{k_i+\phi_x}$ is the Bloch Hamiltonian with the flux parameter $\boldsymbol{\phi}=\phi_x$ introduced. For our 1D stub lattice in the periodic case ($\delta=0$), $\mathcal{H}^0_{k_i+\phi_x}$ is a $3 \times 3$ matrix. Therefore, the full momentum-space Hamiltonian $\mathcal{H}^{\text{full}}_{\text{ks}}(\boldsymbol{\phi})$ is a $3L \times 3L$ block-diagonal matrix, with $\mathcal{H}^0_{k_i+\phi_x}$ occupying the $i$-th $3\times3$ diagonal block. Therefore, the eigenvector matrix $\mathcal{U}(\boldsymbol{\phi})$ corresponding to $\mathcal{H}^{\text{full}}_{\text{ks}}(\boldsymbol{\phi})$ can be written as
\begin{align}\label{eqUphi}
\mathcal{U}(\boldsymbol{\phi})=[U_-(\boldsymbol{\phi})\ U_{\rm f}(\boldsymbol{\phi})\ U_+(\boldsymbol{\phi})],\ U_s(\boldsymbol{\phi})=[|e_1\rangle\otimes|u_s(k_1+\phi_x)\rangle, \cdots, |e_L\rangle\otimes|u_s(k_L+\phi_x)\rangle],
\end{align}
where $U_s(\phi_x)$ is a $3L \times L$ matrix representing the eigenvector matrix corresponding to the $s$-energy sector; $s=-,\ {\rm f},\ +$ correspond to the positive-energy, negative-energy, and flat zero-energy sectors, respectively; $|e_i\rangle$ is an $L \times 1$ column vector with the $i$-th entry equal to 1 and all other entries equal to 0, $|u_s(k_i+\phi_x)\rangle$ is a $3 \times 1$ eigenvector corresponding to the $s$-energy sector of the Bloch Hamiltonian $\mathcal{H}^0_{k_i+\phi_x}$. 

Next, the quantum distance matrix $M_s^{\rm ks}(\boldsymbol{\phi})$, which characterizes the distance between the eigenvector subspaces $U_s(0)$ and $U_s(\boldsymbol{\phi})$ of the full momentum-space Hamiltonian $\mathcal{H}^{\text{full}}_{\text{ks}}(\boldsymbol{\phi})$ as $\phi_x$ varies, is defined as
\begin{align}\label{eqMks}
M^{\mathrm{ks}}_s(\boldsymbol{\phi}) = I - U_s^\dagger(0) U_s(\boldsymbol{\phi})U^\dagger_s(\boldsymbol{\phi}) U_s(0).
\end{align}
Specifically, $M^{\mathrm{ks}}_s(\boldsymbol{\phi})$ is a diagonal matrix, with its $i$-th diagonal element given by $d^{\rm ks}_{s,i}(\boldsymbol{\phi}) = 1 - |\langle u_s(k_i) | u_s(k_i+\phi_x) \rangle|^2$. Therefore, the momentum-space QM $g_{s,xx}(k_i)$ can also be defined in the same form as Eq.~(\ref{eqgni})
\begin{align}\label{eqgksni}
g_{s,xx}(k_i)\equiv g_{s,xx,i}^{\rm FS}=\left.\frac{d^2}{d\phi_x d\phi_x}d^{\rm ks}_{s,i}(\boldsymbol{\phi})\right|_{\phi_x=0}.
\end{align}
Since $M^{\mathrm{ks}}_s(\boldsymbol{\phi})$ is a diagonal matrix, its $i$-th diagonal element $d^{\rm ks}_{s,i}(\boldsymbol{\phi})$ also serves as its $i$-th eigenvalue. Thus, momentum-space QM can be interpreted as a flux-space QM that characterizes the variation of the eigenvalues of the quantum distance matrix. In the periodic case ($\delta = 0$), the real-space Hamiltonian $\mathcal{H}^0_{\text{rs}}(\boldsymbol{\phi})$ and the full momentum-space Hamiltonian $\mathcal{H}^{\text{full}}_{\text{ks}}(\boldsymbol{\phi})$ are unitarily equivalent and share exactly the same eigenvalues. As a result, the momentum-space QM defined in Eq.~(\ref{eqgksni}) is equivalent to the flux-space QM in Eq.~(\ref{eqgni}), yielding identical results. Since the structure of the flux-space QM in Eq.~~(\ref{eqgksni}) closely related to that of the momentum-space QM given on the second line of Eq.~(\ref{eqgkBure}), this suggests that if other equivalent formulations of the momentum-space QM can be identified, then corresponding equivalent forms of the flux-space QM may likewise be constructed. 

From the definition of the momentum-space QM ($D^2=1-|\langle u_{s\boldsymbol{k}}|u_{s\boldsymbol{k}+d\boldsymbol{k}} \rangle|^2\approx \sum_{i,j}\frac{1}{2}g_{s,\alpha\beta}(\boldsymbol{k})dk_\alpha dk_\beta$), the diagonal components of quantum metric tensor can also be expressed in the following limiting form
\begin{align}\label{eqgklim}
\begin{split}
&g_{s,\alpha\alpha}(\boldsymbol{k})\approx \lim_{\delta\phi_{\alpha}\rightarrow 0}\frac{2}{\delta\phi_\alpha \delta\phi_\alpha}[1-\langle u_s(\boldsymbol{k})|u_s(\boldsymbol{k}+(\phi_\alpha,0))\rangle\langle u_s(\boldsymbol{k}+(\phi_\alpha,0))|u_s(\boldsymbol{k})\rangle],\\
&g_{s,\alpha\alpha}(\boldsymbol{k})+g_{s,\beta\beta}(\boldsymbol{k})+2g_{s,\alpha\beta}(\boldsymbol{k})\approx \lim_{\delta\phi_{\alpha,\beta}\rightarrow 0}\frac{2}{\delta\phi_\alpha \delta\phi_\beta}[1-\langle u_s(\boldsymbol{k})|u_s(\boldsymbol{k}+(\phi_\alpha,\phi_\beta))\rangle\langle u_s(\boldsymbol{k}+(\phi_\alpha,\phi_\beta))|u_s(\boldsymbol{k})\rangle],
\end{split}
\end{align}
Eq~(\ref{eqgklim}) can then be equivalently expressed in an alternative form in terms of the flux-space QM associated with the full momentum-space Hamiltonian $\mathcal{H}^{\rm full}_{\rm ks}(\boldsymbol{\phi})$,
\begin{align}\label{eqgFSkilim}
\begin{split}
&g^{\rm FS}_{s,\alpha\alpha}(\boldsymbol{k}_i)\equiv g_{s,\alpha\alpha,i}^{\rm FS}=\lim_{\delta\phi_{\alpha,\alpha\rightarrow 0}}\frac{2}{\delta\phi_\alpha \delta\phi_\alpha}d^{\rm ks}_{s,i}(\boldsymbol{\phi}=(\phi_\alpha,0)),\\
&g_{s,\alpha\alpha,i}^{\rm FS}+g_{s,\beta\beta,i}^{\rm FS}+2g_{s,\alpha\beta,i}^{\rm FS}=\lim_{\delta\phi_{\alpha,\beta\rightarrow 0}}\frac{2}{\delta\phi_\alpha \delta\phi_\beta}d^{\rm ks}_{s,i}(\boldsymbol{\phi}=(\phi_\alpha,\phi_\beta)),\\
&g_{s,\alpha\beta,i}^{\rm FS}=\lim_{\delta\phi_{\alpha,\beta\rightarrow 0}}\frac{1}{\delta\phi_\alpha \delta\phi_\beta}[M^{\rm ks}_s(\boldsymbol{\phi}=(\phi_\alpha,\phi_\beta))-M^{\rm ks}_s(\boldsymbol{\phi}=(\phi_\alpha,0))-M^{\rm ks}_s(\boldsymbol{\phi}=(0,\phi_\beta)]_{i}
\end{split}
\end{align}
where $d^{\rm ks}_{s,i}(\boldsymbol{\phi})$ denotes the $i$-th eigenvalue of the quantum distance matrix $M^{\mathrm{ks}}_s(\boldsymbol{\phi})$, 
 and $[M^{\rm ks}_s(\boldsymbol{\phi}=(\phi_\alpha,\phi_\beta))-M^{\rm ks}_s(\boldsymbol{\phi}=(\phi_\alpha,0))-M^{\rm ks}_s(\boldsymbol{\phi}=(0,\phi_\beta)]_{i}$ denotes the $i$-th eigenvalue of matrix $[M^{\rm ks}_s(\boldsymbol{\phi}=(\phi_\alpha,\phi_\beta))-M^{\rm ks}_s(\boldsymbol{\phi}=(\phi_\alpha,0))-M^{\rm ks}_s(\boldsymbol{\phi}=(0,\phi_\beta)]$. Therefore, Eq. ~(\ref{eqgni}) can also be equivalently expressed in the following limiting form
 \begin{align}\label{eqgFSsilim}
\begin{split}
&g_{s,\alpha\alpha,i}^{\rm FS}=\lim_{\delta\phi_{\alpha,\alpha\rightarrow 0}}\frac{2}{\delta\phi_\alpha \delta\phi_\alpha}d^{\rm rs}_{s,i}(\boldsymbol{\phi}=(\phi_\alpha,0)),\\
&g_{s,\alpha\beta,i}^{\rm FS}=\lim_{\delta\phi_{\alpha,\beta\rightarrow 0}}\frac{1}{\delta\phi_\alpha \delta\phi_\beta}\left[M^{\rm rs}_s(\boldsymbol{\phi}=(\phi_\alpha,\phi_\beta))-M^{\rm rs}_s(\boldsymbol{\phi}=(\phi_\alpha,0))-M^{\rm rs}_s(\boldsymbol{\phi}=(0,\phi_\beta)\right]_{i},
\end{split}
\end{align}
where $d^{\rm rs}_{s,i}(\boldsymbol{\phi})$ denotes the $i$-th eigenvalue of the quantum distance matrix $M^{\mathrm{rs}}_s(\boldsymbol{\phi})$, 
 and $[M^{\rm rs}_s(\boldsymbol{\phi}=(\phi_\alpha,\phi_\beta))-M^{\rm rs}_s(\boldsymbol{\phi}=(\phi_\alpha,0))-M^{\rm rs}_s(\boldsymbol{\phi}=(0,\phi_\beta)]_{i}$ denotes the $i$-th eigenvalue of matrix $[M^{\rm rs}_s(\boldsymbol{\phi}=(\phi_\alpha,\phi_\beta))-M^{\rm rs}_s(\boldsymbol{\phi}=(\phi_\alpha,0))-M^{\rm rs}_s(\boldsymbol{\phi}=(0,\phi_\beta)]$. Eq.~(\ref{eqgFSsilim}) provides a numerically more convenient way to compute the flux-space QM, as it does not require taking derivatives of the eigenvalues, thereby avoiding possible errors caused by eigenvalue ordering in numerical calculations.

\section{Non-interacting Hamiltonian and quantum metric}
\begin{figure}[hbpt]
  \includegraphics[width=8.8cm]{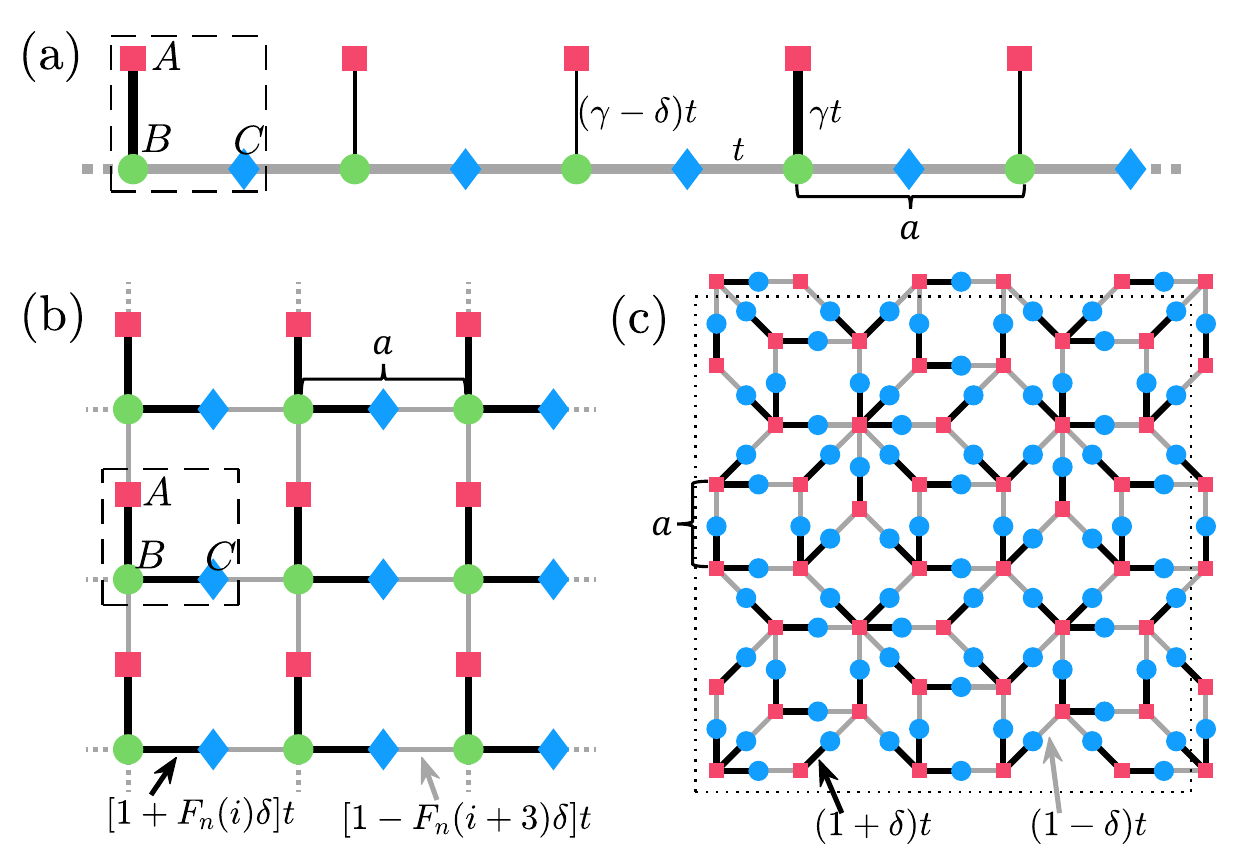}
  \caption{
  (a) Schematic of a 1D Stub-QC of size \( L = 5 \), with the vertical bonds modulated according to the Fibonacci sequence \( F_4 = 01101 \). The three sites in each unit cell (dashed square) are labeled as \( \alpha = A, B, C \). The thin and thick vertical bonds have hopping amplitudes \( (\gamma-\delta) t \) and \( \gamma t \), respectively, which are quasiperiodically distributed according to $F_4$. The coordinate of the leftmost B site is set to \( x_B = 1 \). (b) Schematic of a 2D Lieb-QC composed of three sublattices. The coordinates of the bottom-left B site are set to \( (x_B, y_B) = (1, 1) \).
(c) Schematic of an 2D AB-QC composed of two sublattices (red: A, blue: B), featuring two hopping strengths: black bonds correspond to a hopping amplitude of \( 1+\delta \), and gray bonds correspond to a hopping amplitude of \( 1-\delta \). In (a) and (b), the CBC connects the boundaries through the dotted lines along the edges, while in (c), the lattice sites on the upper and right boundaries outside the black dashed box are treated as periodic replicas of the lower and left boundaries. Bonds crossing the dashed box represent the CBC connections between the lower and left edges. In (a), (b), and (c), \( t = 1 \) and the lattice spacing \( a = 1 \) are set.
}\label{figS0}
\end{figure}

\begin{figure}[hbpt]
  \includegraphics[width=10cm]{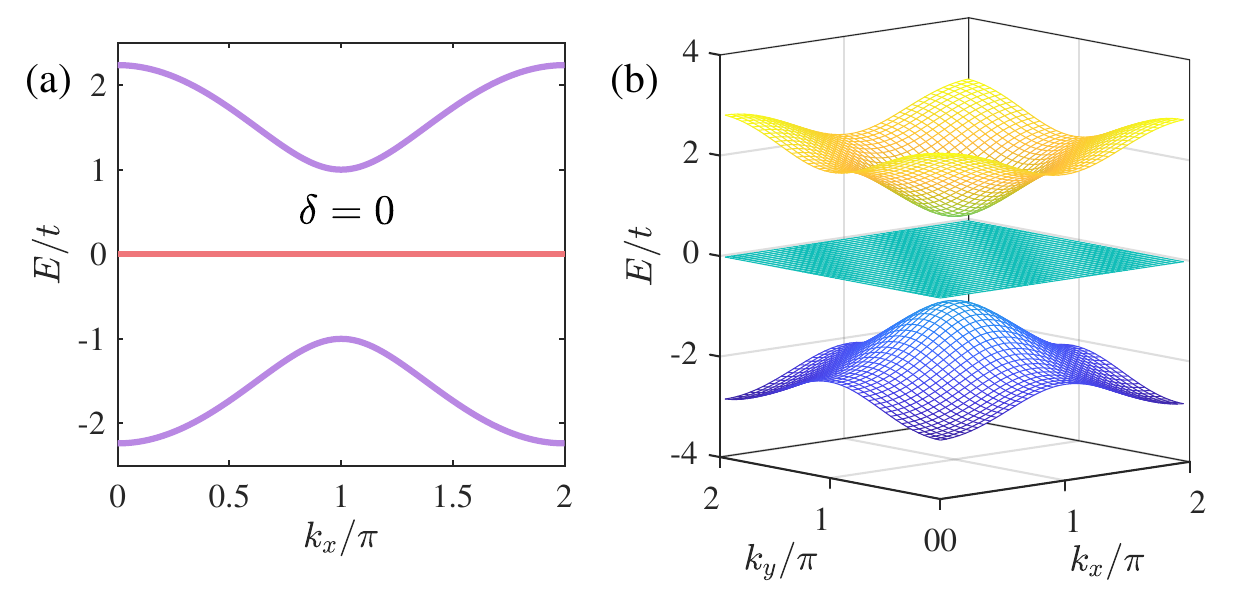}
  \caption{Band structures of the periodic stub lattice with $\gamma=1$ (a) and the Lieb lattice with periodic staggered hopping at $\delta=0.3$ (b).}\label{figS1}
\end{figure}

Stub and Lieb quasicrystals are constructed by introducing Fibonacci quasiperiodic modulation to the hopping amplitudes. For the 1D stub quasicrystal (Stub-QC), while the horizontal hopping is uniform with $t_{ij}^{BC} = t$, the vertical hopping $t_{ij}^{AB}$ is quasiperiodically modulated by the Fibonacci sequence as $t_{ij}^{AB} = [\gamma - F_n(x_B)\delta] t$, where $\delta$ controls the modulation strength, \(x_B\) denotes the integer \(x\)-coordinate of the $B$ site and $F_n(x_B)$ is the $x_B$-th number in the Fibonacci sequence \(F_n\)~\cite{Fn}. Since each $F_n(x_B)$ is either 0 or 1, the vertical hopping $t_{ij}^{AB}$ takes values $\gamma t$ or $(\gamma - \delta)t$. 
For the 2D Lieb quasicrystal (Lieb-QC), hoppings along each row and column are modulated by the same Fibonacci sequence: intracell hoppings are \(t_{ij}^{BC}=[1+F_n(2x_B-1)\delta]t,\ t_{ij}^{BA}=[1+F_n(2y_B-1)\delta]t\), while the intercell hoppings to the right and upward neighboring sites are \(t_{ij}^{CB}=[1-F_n(2x_B)\delta]t,\ t_{ij}^{AB}=[1-F_n(2y_B)\delta]t\), where $(x_B,y_B)$ labels the integer coordinate components of the $B$ site.

For the 1D Stub-QC, without the Fibonacci modulation (\(\delta = 0\)), the system recovers the periodicity, and the non-interacting Hamiltonian in the momentum-space can be expressed as $H_{0}=\sum_{\boldsymbol{k},\sigma}\hat{c}^\dagger_{\boldsymbol{k}\sigma}\mathcal{H}^0_{\boldsymbol{k}}\hat{c}_{\boldsymbol{k}\sigma}$ with the basis $\hat{c}_{\boldsymbol{k}\sigma}=(\hat{c}^A_{\boldsymbol{k}\sigma},\ \hat{c}^B_{\boldsymbol{k}\sigma},\ \hat{c}^C_{\boldsymbol{k}\sigma})^{T}$ and
\begin{align}\label{}
\mathcal{H}^0_{\boldsymbol{k}}=\left[\begin{array}{ccc}
0 & \gamma t & 0 \\
\gamma t & 0 & f_{\boldsymbol{k}}  \\
0 & f_{\boldsymbol{k}}^* & 0
\end{array}\right],
\end{align}
where $f_{\boldsymbol{k}}=t[e^{-ik_x/2}+e^{ik_x/2}]$ (In this calculation, we set the distance between the $B$ and $C$ sites is $1/2$ within the unit cell of the size $a=1$, $\sigma = \uparrow,\downarrow$ is the spin index and $\hat{c}^\alpha_{\boldsymbol{k}\sigma}=(1/{\sqrt{L}})\sum_ie^{i\boldsymbol{k}\cdot \boldsymbol{r}_{i\alpha}}\hat{c}_{i\alpha,\sigma}$. By solving the eigenvalue problem \( {\cal H}^0_{\boldsymbol{k}} |u_{s\boldsymbol{k}}\rangle = E_{s\boldsymbol,{\boldsymbol{k}}} |u_{s\boldsymbol{k}}\rangle \), we directly obtain the Bloch functions $|u_{s\boldsymbol{k}}\rangle$ and the energy spectrum $E_{s,\boldsymbol{k}}$, which consists of one flat band $E_{{\rm f},\boldsymbol{k}} = 0$ and two dispersive bands $E_{\pm,\boldsymbol{k}}=\pm t\sqrt{\gamma^2+4\cos^2(k_x/2)}$ [see Fig.~\ref{figS1}(a)]. 
The eigenstate with zero-energy is spanned by compact localized states (CLSs)~\cite{PhysRevB.34.5208,PhysRevB.95.115135,PhysRevB.99.045107} formed near each unit cell $i$, given by $|{\rm CLS}\rangle_i = \frac{1}{\sqrt{\gamma^2+2}}(|A_i\rangle+|A_{i+1}\rangle-\gamma|C_i\rangle)$ with $|\alpha_{i}\rangle=\hat{c}^\dagger_{i\alpha}|0\rangle$ the local basis on the $\alpha$ site of the $i$-th unit cell.

The momentum-space QM(Fubini-Study metric) of a selected band $s$, $g_{s,\alpha\beta}(\boldsymbol{k})$, corresponds to the real part of the quantum geometric tensor: 
\begin{align}\label{QMk}
g_{s,\alpha\beta}(\boldsymbol{k})=2{\rm Re}\{\langle \partial_{k_\alpha} u_{s\boldsymbol{k}}|(1-|u_{s\boldsymbol{k}}\rangle\langle u_{s\boldsymbol{k}}|)|\partial_{k_\beta} u_{s\boldsymbol{k}}\rangle)\}.
\end{align}
The QM for the flat band is 
\begin{align}\label{eqstubgk}
g_{{\rm f},xx}(k_x)=\frac{\gamma^2\sin^2\frac{k_x}{2}}{\left(\gamma^2+4\cos^2\frac{k_x}{2}\right)^2},
\end{align}
the analytical integrated quantum metric (IQM) $\mathcal{Q}_{{\rm f},xx}$ associated with the flat band can be obtained as follows 
\begin{align}\label{}
\mathcal{Q}_{{\rm f},xx}=\frac{1}{2\pi}\int_{\rm BZ} g_{{\rm f},xx}(k_x)dk_x=\frac{1}{L}\sum_{k_x} g_{{\rm f},xx}(k_x).
\end{align}

For the 2D Lieb-QC, the quasiperiodic modulation can be removed by setting all elements of the Fibonacci sequence to \( F_n(i) = 1 \), which reduces the system to a 2D periodic Lieb lattice with staggered hopping. The non-interacting Hamiltonian in the momentum-space can be expressed as $H_{0}=\sum_{\boldsymbol{k},\sigma}\hat{c}^\dagger_{\boldsymbol{k}\sigma}\mathcal{H}^0_{\boldsymbol{k}}\hat{c}_{\boldsymbol{k}\sigma}$ with the basis $\hat{c}_{\boldsymbol{k}\sigma}=(\hat{c}^A_{\boldsymbol{k}\sigma},\ \hat{c}^B_{\boldsymbol{k}\sigma},\ \hat{c}^C_{\boldsymbol{k}\sigma})^{T}$ and
\begin{align}\label{}
\mathcal{H}^0_{\boldsymbol{k}}=2t\left[\begin{array}{ccc}
0 & a^*_{\boldsymbol{k}} & 0 \\
a_{\boldsymbol{k}} & 0 & b_{\boldsymbol{k}}  \\
0 & b_{\boldsymbol{k}}^* & 0
\end{array}\right],
\end{align}
where $a_{\boldsymbol{k}}=\cos(k_y/2)+i\delta\sin(k_y/2)$, $b_{\boldsymbol{k}}=\cos(k_x/2)+i\delta\sin(k_x/2)$, $\sigma = \uparrow,\downarrow$ is the spin index. By solving the eigenvalue problem \( {\cal H}^0_{\boldsymbol{k}} |u_{s\boldsymbol{k}}\rangle = E_{s\boldsymbol,{\boldsymbol{k}}} |u_{s\boldsymbol{k}}\rangle \), we directly obtain the Bloch functions $|u_{s\boldsymbol{k}}\rangle$ and the energy spectrum $E_{s,\boldsymbol{k}}$, which consists of one flat band $E_{{\rm f},\boldsymbol{k}} = 0$ and two dispersive bands \\
$E_{\pm,\boldsymbol{k}}=\pm 2t\sqrt{1+\delta^2+(1-\delta^2)(\cos k_x+\cos k_y)/2}$ [see Fig.~\ref{figS1}(b)]. 

The QM for the flat band is 
\begin{align}\label{eqLiebgk}
\begin{split}
g_{{\rm f},\alpha\alpha}(\boldsymbol{k})&=\frac{[1+\delta^2+(\delta^2-1)\cos k_\alpha][1+\delta^2-(\delta^2-1)\cos k_\beta]}{2[2(1+\delta^2)+(1-\delta^2)(\cos k_\alpha+ \cos k_\beta)]^2},\\
g_{{\rm f},\alpha\beta}(\boldsymbol{k})&={\rm Re}\frac{[2i\delta-(\delta^2-1)\sin k_\alpha][2i\delta+(\delta^2-1)\sin k_\beta]}{2[2(1+\delta^2)+(1-\delta^2)(\cos k_\alpha+ \cos k_\beta)]^2}.
\end{split}
\end{align}
It should be noted that the overall sign of the off-diagonal component of the quantum metric \( g_{{\rm f},\alpha\beta}(\boldsymbol{k}) \) depends on the choice of coordinate orientation. If the direction of either coordinate axis is reversed, \( g_{{\rm f},\alpha\beta}(\boldsymbol{k}) \) acquires an additional negative sign. Then the analytical integrated quantum metric (IQM) $\mathcal{Q}_{{\rm f},\alpha\beta}$ associated with the flat band can be obtained as follows 
\begin{align}\label{}
\mathcal{Q}_{{\rm f},\alpha\beta}=\frac{1}{2\pi}\int_{\rm BZ} g_{{\rm f},\alpha\beta}(\boldsymbol{k})d\boldsymbol{k}=\frac{2\pi}{L_xL_y}\sum_{\boldsymbol{k}} g_{{\rm f},\alpha\beta}(\boldsymbol{k}).
\end{align}

\begin{figure}[hbpt]
  \includegraphics[width=16cm]{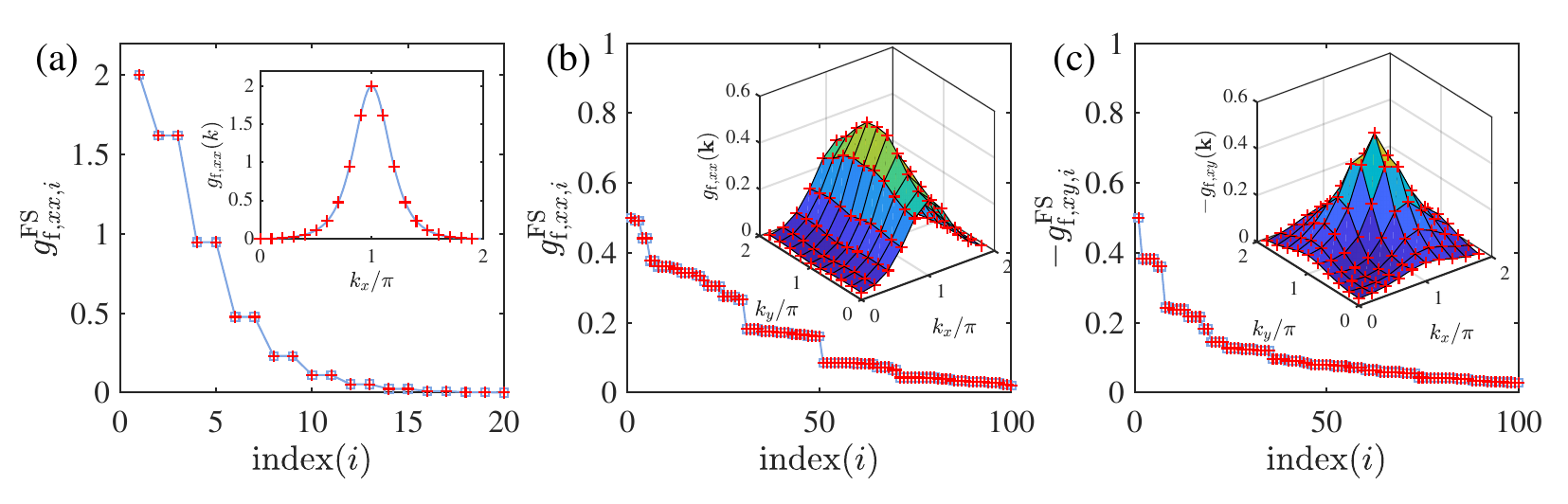}
  \caption{In panels (a), (b), and (c), the blue lines with square markers correspond to \( g^{\rm FS}_{{\rm f},xx,i} \) of the stub lattice, \( g^{\rm FS}_{{\rm f},xx,i} \) of the Lieb lattice, and \( -g^{\rm FS}_{{\rm f},xy,i} \) of the Lieb lattice, respectively. The red plus markers represent the values of the quantum metric at discrete momentum points in momentum space (as indicated by the red plus markers in the insets), sorted in descending order. The inset in panel (a) shows the results for the periodic stub lattice corresponding to Eq.~(\ref{eqstubgk}) at \(\gamma = 1\), with lattice size \(L=L_x = 20\). The insets in panels (b) and (c) show the results for the periodic Lieb lattice with staggered hopping corresponding to Eq.~(\ref{eqLiebgk}) at \(\delta = 0.5\), with lattice size \(L_x = L_y = 10\).
}\label{figS2}
\end{figure}

Here, we show that for finite-size periodic stub lattices and Lieb lattices with staggered hopping, the momentum-space QM evaluated at discrete momentum points is equivalent to the flux-space QM formulated in the real-space representation [Eq.~(\ref{eqgni}) or Eq.~(\ref{eqgFSsilim})], as illustrated in Fig.~\ref{figS2}.

\begin{figure}[hbpt]
  \includegraphics[width=14cm]{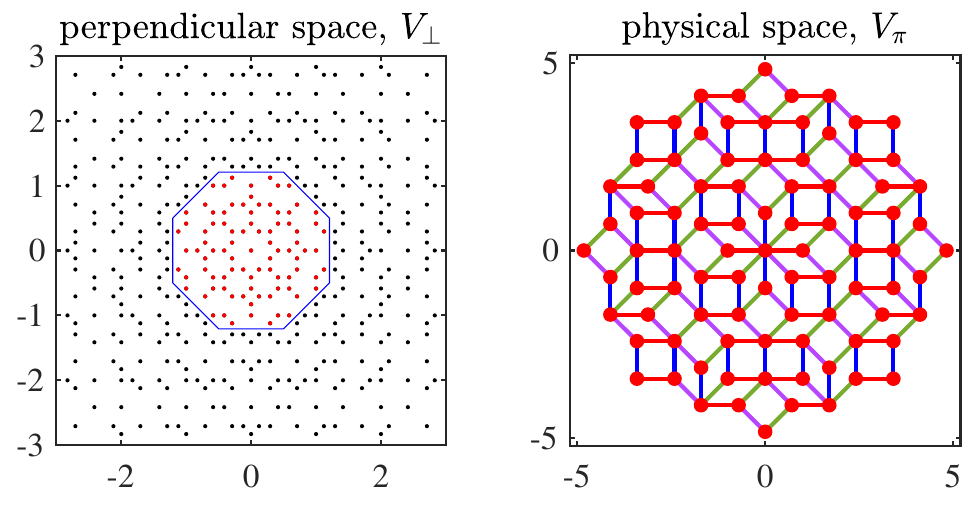}
  \caption{Left panel: Projection of the lattice points of a hypercubic lattice with size \( L = 5 \) (i.e., \( x_{\alpha=1,2,3,4} = -2, -1, 0, 1, 2 \) onto the perpendicular space. The blue regular octagon represents the selection window with a side length of 1, and the accepted points within the window are marked in red. Right panel: The projections onto the physical space of the four-dimensional lattice points corresponding to the red-marked points in the left panel. The four colors of the bonds represent the projections of the four types of nearest-neighbor bond vectors of the hypercubic lattice onto the physical space: red for \( \boldsymbol{w}_1 \), blue for \( \boldsymbol{w}_2 \), green for \( \boldsymbol{w}_3 \), and purple for \( \boldsymbol{w}_4 \).
}\label{figS_AB}
\end{figure}

We know that the Ammann-Beenker tiling quasicrystal can be obtained by projecting a four-dimensional hypercubic lattice onto a two-dimensional space~\cite{https://doi.org/10.1002/ijch.202300119,jeon2022discovery,li2025unconventionalalter}. Let \(\{\boldsymbol{e}_1, \boldsymbol{e}_2, \boldsymbol{e}_3, \boldsymbol{e}_4\}\) (\(\boldsymbol{e}_1=(1,0,0,0)^T\), \(\boldsymbol{e}_2=(0,1,0,0)^T\), \(\boldsymbol{e}_3=(0,0,1,0)^T\), and \(\boldsymbol{e}_4=(0,0,0,1)^T\)) be the basis vectors of the four-dimensional hypercubic lattice. Then, the position of any lattice site on the hypercubic lattice can be expressed as $R = \sum_{i=1}^4 x_i \boldsymbol{e}_i$,
and can thus be labeled by the four-dimensional coordinate \(X=(x_1, x_2, x_3, x_4)^T\). Then, by applying two projection matrices, 
\begin{align}\label{s11}
\begin{split}
\mathcal{S}_\pi=\left(\begin{array}{cccc}
1 & 0 & \frac{1}{\sqrt{2}} & -\frac{1}{\sqrt{2}} \\
0 & 1 & \frac{1}{\sqrt{2}} & \frac{1}{\sqrt{2}}\\
\end{array}\right),\ \ 
\mathcal{S}_\perp=\left(\begin{array}{cccc}
1 & 0 & -\frac{1}{\sqrt{2}} & \frac{1}{\sqrt{2}} \\
0 & 1 & -\frac{1}{\sqrt{2}} & -\frac{1}{\sqrt{2}} \\
\end{array}\right),
\end{split}
\end{align}
the hypercubic lattice sites can be projected onto the physical space ($V_\pi$) and the perpendicular space ($V_\perp$), respectively. The Ammann-Beenker lattice structure in physical space is determined by the selection window (or acceptance window) in the perpendicular space. We choose a regular octagon centered at the origin with a side length of 1 as the selection window in the perpendicular space. For lattice points whose perpendicular projections ($\mathcal{S_\perp}X$) fall within this window, we keep their corresponding projections ($\mathcal{S}_\pi X$) in the physical space, while those whose perpendicular projections lie outside the window are discarded. In this way, we obtain the standard Ammann-Beenker quasicrystal (see Fig.~\ref{figS_AB}). The four nearest-neighbor hopping bonds of the hypercubic lattice, corresponding to the four vectors \(\boldsymbol{e}_1\), \(\boldsymbol{e}_2\), \(\boldsymbol{e}_3\), and \(\boldsymbol{e}_4\), are projected onto the physical space to form the four bond vectors of the Ammann-Beenker quasicrystal,
\begin{align}\label{s11}
\begin{split}
\boldsymbol{w}_1=\mathcal{S}_\pi\boldsymbol{e}_1=(1,0)^T,\ \boldsymbol{w}_2=\mathcal{S}_\pi\boldsymbol{e}_2=(0,1)^T,\ 
\boldsymbol{w}_3=\mathcal{S}_\pi\boldsymbol{e}_3=(\frac{1}{\sqrt{2}},\frac{1}{\sqrt{2}})^T,\ 
\boldsymbol{w}_4=\mathcal{S}_\pi\boldsymbol{e}_4=(-\frac{1}{\sqrt{2}},\frac{1}{\sqrt{2}})^T.
\end{split}
\end{align}

\begin{figure}[hbpt]
  \includegraphics[width=14cm]{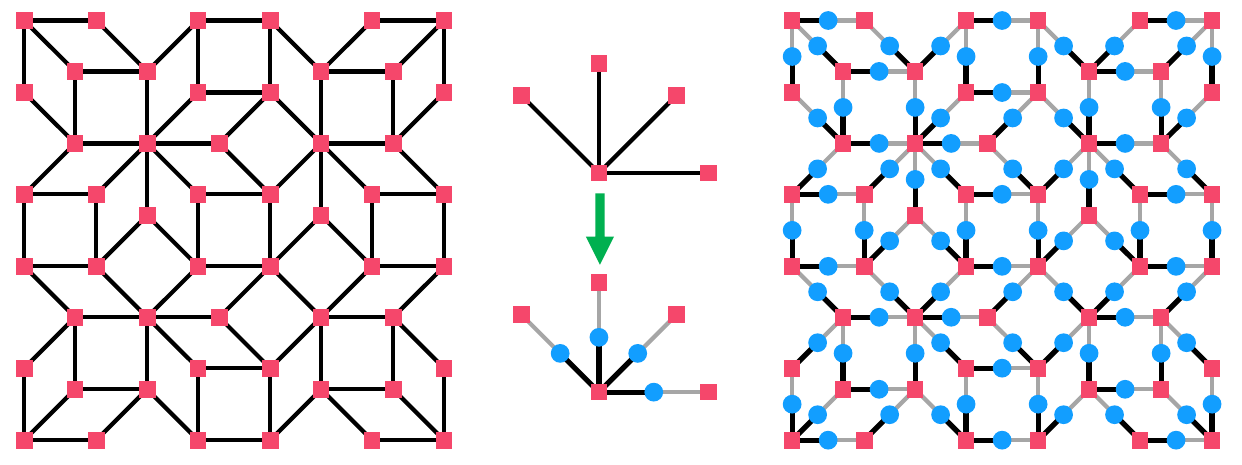}
  \caption{Left: Ammann-Beenker tiling quasicrystal constructed using the deflation/inflation method. Middle: bond substitution process along the four directional vectors. Right: our Ammann-Beenker-type quasicrystal (AB-QC) obtained by applying the bond substitution shown in the middle panel to the structure on the left.}\label{figS_ABsub}
\end{figure}
The Ammann-Beenker-type quasicrystal (AB-QC) shown in Fig.~\ref{figS0} (c) can also be obtained by applying the above projection method to a four-dimensional Lieb lattice with staggered hopping. The four-dimensional Lieb lattice can be constructed straightforwardly as follows. We first define the lattice sites of the original hypercubic lattice as the A sublattice, and then add a new set of B sublattice sites at the centers of all bonds of the hypercubic lattice. Let the coordinates of the A sublattice be defined as \( X_A = (x_1, x_2, x_3, x_4) \), where each \( x_{\alpha=1,2,3,4} \) is an integer. Then, the coordinates of the B sublattice are given by \( X_B = (x_1+\frac{1}{2}, x_2, x_3, x_4),\ (x_1, x_2+\frac{1}{2}, x_3, x_4),\  (x_1, x_2, x_3+\frac{1}{2}, x_4),\ (x_1, x_2, x_3, x_4+\frac{1}{2})\). After adding the B sublattice, each original bond of the hypercubic lattice is divided into two bonds, forming periodically staggered hoppings of \(1 + \delta\) and \(1 - \delta\) along each dimension. The lattice structure shown in Fig.~\ref{figS0}(c) can then be obtained through the following steps. First, for the A sublattice, we retain only those A sites whose perpendicular projections fall within the selection window. For the B sublattice, we retain only those B sites that are connected to the selected A sites. Finally, these selected A and B sites, together with their connectivity, are projected onto the physical space according to the connection structure defined in the four-dimensional Lieb lattice.  

In fact, from the projection procedure of the four-dimensional Lieb lattice described above, it is clear that our Ammann-Beenker-type quasicrystal shown in Fig. ~\ref{figS0}(c) can be directly obtained by performing a bond substitution along the four bond directions (as shown by the substitution process indicated by the green arrows in the middle panel of Fig.~\ref{figS_ABsub}.) of the original Ammann-Beenker lattice. To facilitate the generation of a lattice structure with a rectangular outline suitable for applying CBC, we adopt an alternative deflation/inflation method~\cite{https://doi.org/10.1002/ijch.202300119,PhysRevB.102.115125,PhysRevB.104.144511} to construct the Ammann-Beenker tiling quasicrystal. By subsequently performing bond substitution, we can conveniently obtain our AB-QC, as illustrated in Fig.~\ref{figS_ABsub}.

\begin{figure}[hbpt]
  \includegraphics[width=16cm]{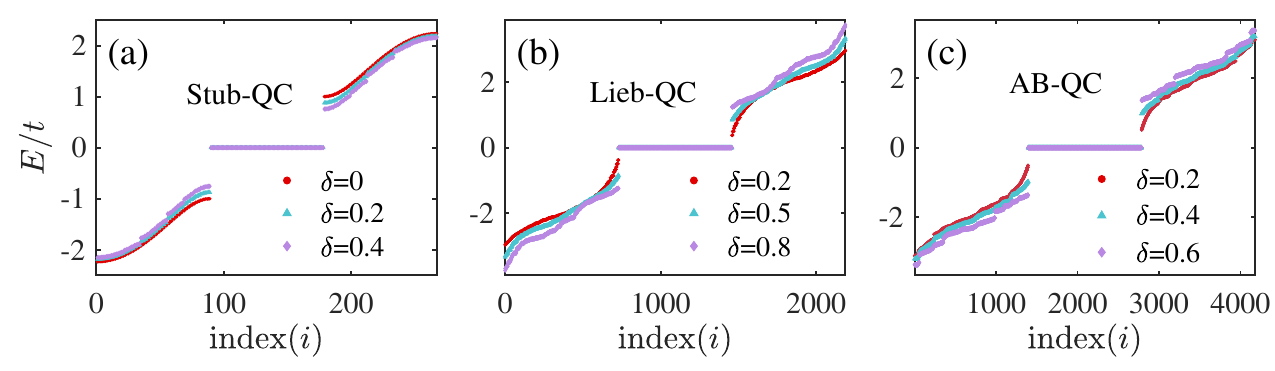}
  \caption{Non-interacting eigen-energy spectra of the three quasicrystals (under CBC conditions) for (a) Stub-QC, (b) Lieb-QC, and (c) AB-QC.}\label{figS_En}
\end{figure}

In the quasiperiodic case, for the three types of quasicrystals, we can directly construct the Hamiltonian in the real-space representation, $H_0 = \sum_\sigma \hat{C}_\sigma^\dagger \mathcal{H}^0_{\rm rs} \hat{C}_\sigma$, where $\hat{C}_{\sigma}=(\hat{c}_{1,\sigma},\hat{c}_{2,\sigma},\dots,\hat{c}_{N_t,\sigma})^T$ ($N_t$ is the total number of sites), and diagonalize the real-space Hamiltonian matrix \( \mathcal{H}^0_{\rm rs} \) to obtain their eigenenergy spectrum. As shown in Fig.~\ref{figS_En}, the non-interacting spectra of the three quasicrystals all exhibit a flat-energy sector at zero energy, together with positive- and negative-energy sectors. The flat-energy sector remains unchanged and is independent of $\delta$, while the non-flat-energy sector evolves with $\delta$. For the 1D stub-QC, the energy gap between the flat-energy and non-flat-energy sectors decreases as $\delta$ increases, whereas for the 2D Lieb-QC and AB-QC, the gap between the flat-energy and non-flat-energy sectors increases with increasing $\delta$.

\section{Compact localized states in the three quasicrystals}
The non-interacting Hamiltonians on the three quasicrystals can be generally written as follows
\begin{align}\label{s11}
\begin{split}
\hat{H_0}=&\sum_{\langle i,j\rangle}(t^{\xi_i \eta_j}_{ij}c_{i}^\dagger c_{j}+{\rm H.c.} ),
\end{split}
\end{align}
where we omit the spin index, as it takes the same form for both spin-up and spin-down electrons.

For Stub-QC, in real space, the above Hamiltonian can be expressed as $\hat{H}_{0}=\hat{C}^\dagger\mathcal{H}^0_{\rm rs}\hat{C}$, where the basis is $\hat{C}=(\hat{c}_1^{A},\hat{c}_1^{B},\hat{c}_1^{C},\dots,\hat{c}_L^{A},\hat{c}_L^{B},\hat{c}_L^{C})^T$. Here, we use \(\hat{c}_x^{\xi}\) to denote the original operator \(\hat{c}_i\), where $x$ represents the position of the unit cell, equivalent to the coordinate of the B site, and $\xi$ labels the sublattice. Assuming the eigenfunction takes the form \( |\Psi\rangle=\sum_{x=1}^L(a_x|A_x\rangle+b_x|B_x\rangle+c_x|C_x\rangle) \), the eigenvalue equation is written as \( \mathcal{H}^0_{\rm rs}|\Psi\rangle = E|\Psi\rangle \).
Under closed boundary conditions (CBC) and in the absence of quasiperiodic modulation, the following set of equations must be satisfied for the zero eigenvalue:
\begin{align}\label{s12}
\begin{split}
&\gamma b_x=0,\\
&\gamma a_x+c_x+c_{x-1}=0,\\
&b_x+b_{x+1}=0,
\end{split}
\end{align}
where $x=1,2,\cdots ,L$. When the subscript becomes $0$ or $L+1$, which are outside the index range, they correspond to $x = L$ and $x = 1$, respectively, due to the imposed CBC. From the equations in Eq.~(\ref{s12}), it is observed that \( b_x \) is always equal to zero. The second set of equations in Eq.~(\ref{s12}) leads to the solutions of the compact localized states. Totally, we have \( L \) equations, namely $\gamma a_x+c_x+c_{x-1}=0\ (x=1,2,\cdots ,L)$.  
It is observed that the term $c_{x-1}$ in the $x$-th equation ($\gamma a_x+c_x+c_{x-1}=0$) couples the set of equations.
An solution that decouples the \( x \)-th and \( (x+1) \)-th equations from the other ones, can be found by setting \( c_{x-1} = 0 \) and \( c_{x+1} = 0 \). As a result, a compact localized state (CLS) can be formed from nonzero \( a_x \), \( a_{x+1} \), and \( c_x \): $|{\rm CLS}\rangle_x=\frac{1}{\sqrt{\gamma^2+2}}(|A_x\rangle+|A_{x+1}\rangle-\gamma|C_x\rangle)$.

When the quasiperiodic modulation is present (\( \delta \neq 0 \)), \( b_x \) still equals zero, and the nontrivial equation in the second line of Eq.~(\ref{s12}) for the Stub-QC [see Fig.~\ref{figS0}(a)] becomes
\begin{align}\label{}
\begin{split}
\left[\gamma-F_n(x)\delta\right] a_x+c_x+c_{x-1}=0.
\end{split}
\end{align}
The coefficient in front of \( a_i \) can be either \( \gamma \) or \(\gamma_{\delta}= \gamma - \delta \), depending on the $i$-th element of the Fibonacci sequence. For our chosen Fibonacci sequence, any two adjacent elements can only appear in one of three combinations: '11', '01', or '10'. As a result, the coefficients in front of \( a_i \) and \( a_{i+1} \) in two adjacent equations can take one of the following three forms \(( \gamma_\delta,\ \gamma_\delta) \), \( (\gamma,\ \gamma_\delta) \), or \( (\gamma_\delta,\ \gamma) \). Therefore, we can construct three types of CLSs corresponding to these three cases:
\begin{align}\label{}
\begin{split}
|{\rm CLS}^{(\gamma_\delta,\gamma_\delta)}\rangle_x&=\frac{1}{\sqrt{\gamma_\delta^2+2}}(|A_x\rangle+|A_{x+1}\rangle-\gamma_\delta|C_x\rangle),\\
|{\rm CLS}^{(\gamma,\gamma_\delta)}\rangle_x&=\frac{1}{\sqrt{\gamma_\delta^2+1+(\frac{\gamma_\delta}{\gamma})^2}}(\frac{\gamma_\delta}{\gamma}|A_x\rangle+|A_{x+1}\rangle-\gamma_\delta|C_x\rangle),\\
|{\rm CLS}^{(\gamma_\delta,\gamma)}\rangle_x&=\frac{1}{\sqrt{\gamma_\delta^2+1+(\frac{\gamma_\delta}{\gamma})^2}}(|A_x\rangle+\frac{\gamma_\delta}{\gamma}|A_{x+1}\rangle-\gamma_\delta|C_x\rangle).
\end{split}
\end{align}

For the 2D Lieb-QC, the real-space Hamiltonian can be written as $\hat{H}_{0}=\hat{C}^\dagger\mathcal{H}^0_{\rm rs}\hat{C}$, where the basis is $\hat{C}=(\hat{c}^A_{1,1},\hat{c}^B_{1,1},\hat{c}^C_{1,1},\dots,\hat{c}^A_{L_x,L_y},\hat{c}^B_{L_x,L_y},\hat{c}^C_{L_x,L_y})^T$. Here, we use \(\hat{c}_{x,y}^{\xi}\) to denote the original operator \(\hat{c}_i\), where $x,y$ denote the coordinate components of the unit cell position, equivalent to the coordinate of the B site, and $\xi$ labels the sublattice. Assuming the eigenfunction takes the form \( |\Psi\rangle=\sum_{y=1}^{L_y}\sum_{x=1}^{L_x}(a_{x,y}|A_{x,y}\rangle+b_{x,y}|B_{x,y}\rangle+c_{x,y}|C_{x,y}\rangle) \), the eigenvalue equation is written as \( \mathcal{H}^0_{\rm rs}|\Psi\rangle = E|\Psi\rangle \).
Under closed boundary conditions (CBC), the following set of equations must be satisfied for the zero eigenvalue:
\begin{align}\label{}
\begin{split}
&b_{x,y}=0,\\
&[1+F_n(2x-1)\delta]c_{x,y}+[1+F_n(2x-2)\delta]c_{x-1,y}+(1+F_n(2y-1)\delta)a_{x,y}+(1+F_n(2y-2)\delta)a_{x,y-1}=0.
\end{split}
\end{align}
We can obtain the compact localized state (CLS) localized at the unit cell \((x, y)\) from the following set of equations
\begin{align}\label{}
\begin{split}
&[1+F_n(2x-1)\delta]c_{x,y}+[1-F_n(2x-2)\delta]c_{x-1,y}+[1+F_n(2y-1)\delta]a_{x,y}+[1-F_n(2y-2)\delta]a_{x,y-1}=0,\\
&[1+F_n(2x-1)\delta]c_{x,y+1}+[1-F_n(2x-2)\delta]c_{x-1,y+1}+[1+F_n(2y+1)\delta]a_{x,y+1}+[1-F_n(2y)\delta]a_{x,y}=0,\\
&[1+F_n(2x+1)\delta]c_{x+1,y}+[1-F_n(2x)\delta]c_{x,y}+[1+F_n(2y-1)\delta]a_{x+1,y}+[1-F_n(2y-2)\delta]a_{x+1,y-1}=0,\\
&[1+F_n(2x+1)\delta]c_{x+1,y+1}+[1-F_n(2x)\delta]c_{x,y+1}+[1+F_n(2y+1]\delta)a_{x+1,y+1}+[1-F_n(2y)\delta]a_{x+1,y}=0.
\end{split}
\end{align}
The $|{\rm CLS}\rangle_{x,y}$ is distributed only over the four sites \( |A_{x,y}\rangle, |C_{x,y}\rangle, |A_{x+1,y}\rangle, |C_{x,y+1}\rangle \), i.e., only the amplitudes 
\( a_{x,y}, c_{x,y}, a_{x+1,y}, c_{x,y+1} \) are nonzero. This requires that these amplitudes satisfy
\begin{align}\label{}
\begin{split}
&[1+F_n(2x-1)\delta]c_{x,y}+[1+F_n(2y-1)\delta]a_{x,y}=0,\\
&[1+F_n(2x-1)\delta]c_{x,y+1}+[1-F_n(2y)\delta]a_{x,y}=0,\\
&[1-F_n(2x)\delta]c_{x,y}+[1+F_n(2y-1)\delta]a_{x+1,y}=0,\\
&[1-F_n(2x)\delta]c_{x,y+1}+[1-F_n(2y)\delta]a_{x+1,y}=0.
\end{split}
\end{align}
By setting \(a_{x,y} = 1 + F_n(2x-1)\delta\), one obtains \(a_{x+1,y} = 1 -F_n(2x) \delta,\ c_{x,y}=-[1+F_n(2y-1)\delta],\ c_{x,y+1}=-[1-F_n(2y)\delta]\). Therefore, we obtain
\begin{align}\label{}
\begin{split}
|{\rm CLS}\rangle_{x,y}=\frac{1}{\sqrt{C_N}}\left\{[1 + F_n(2x-1)\delta]|A_{x,y}\rangle+[1 -F_n(2x)\delta]|A_{x+1,y}-[1+F_n(2y-1)\delta]|C_{x,y}\rangle-[1-F_n(2y)\delta]|C_{x,y+1}\rangle\right\},
\end{split}
\end{align}
where $C_N=[1 + F_n(2x-1)\delta]^2+[1 -F_n(2x)\delta]^2+[1+F_n(2y-1)\delta]^2+[1-F_n(2y)\delta]^2$. When the quasiperiodic modulation is removed, i.e., by setting any \(F(i) = 1\), all $|{\rm CLS}\rangle_{x,y}$ take the same form,
\begin{align}\label{}
\begin{split}
|{\rm CLS}\rangle_{x,y}=\frac{(1+\delta)|A_{x,y}\rangle+(1-\delta)|A_{x+1,y}\rangle-(1+\delta)|C_{x,y}\rangle-(1-\delta)|C_{x,y+1}\rangle}{\sqrt{2(1+\delta)^2+2(1-\delta)^2}},
\end{split}
\end{align}
and exhibit translational symmetry.

\begin{figure}[hbpt]
  \includegraphics[width=14cm]{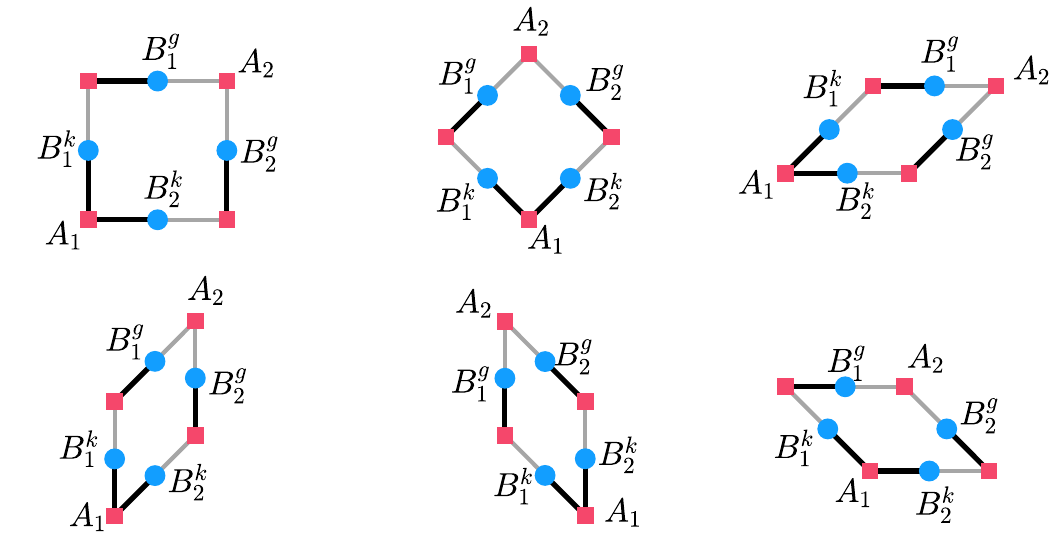}
  \caption{The six types of topologically equivalent square or \(45^\circ\) rhombic tilings contained in the AB-QC in Fig.~\ref{figS0}(c).}\label{figS_ABCLS}
\end{figure}

For the 2D AB-QC, within each square or \(45^{\circ}\) rhombic tile, there exists a CLS localized on the four B sites located at the midpoints of its edges. Each square or \(45^{\circ}\) rhombic tile shares a topologically equivalent structure (see Fig.~\ref{figS_ABCLS}): two of the B sites (labeled \(B_1^k\) and \(B_2^k\) are connected to one A site (labeled \(A_1\)) by black bonds (with hopping amplitude \(1+\delta\))), while the other two B sites (labeled \(B_1^g\) and \(B_2^g\)) are connected to the A site (labeled \(A_2\)), which is diagonally opposite to \(A_1\), by gray bonds (with hopping amplitude \(1-\delta\)).
Assuming the CLS takes the form
$
|{\rm CLS}\rangle =
\frac{1}{\sqrt{b_{1,k}^2+b_{2,k}^2+b_{1,g}^2+b_{2,g}^2}}
\left(b_{1,k}|B_1^k\rangle + b_{2,k}|B_2^k\rangle + b_{1,g}|B_1^g\rangle + b_{2,g}|B_2^g\rangle\right),
$
the following equations must be satisfied,
\begin{align}\label{}
\begin{split}
&(1+\delta)b_{1,k}+(1+\delta)b_{2,k}=0,\ (1-\delta)b_{1,g}+(1-\delta)b_{2,g}=0,\\
&(1-\delta)b_{1,k}+(1+\delta)b_{1,g}=0,\ (1-\delta)b_{2,k}+(1+\delta)b_{2,g}=0.
\end{split}
\end{align}
Thus, we can see that on each square or $45^{\circ}$ rhombic tile, there exists a CLS of the following form
\begin{align}\label{}
\begin{split}
|{\rm CLS}\rangle = \frac{1}{\sqrt{2(1+\delta)^2+2(1-\delta)^2}}\left[(1+\delta)|B_1^k\rangle -(1+\delta)|B_2^k\rangle + (1-\delta)|B_1^g\rangle -(1-\delta)|B_2^g\rangle\right].
\end{split}
\end{align}

\section{The BdG mean-field theory}
We employ the Bogoliubov-de Gennes (BdG) mean-field method~\cite{PhysRevB.65.014501(2001), zhu2016bogoliubov, PhysRevB.101.144503(2020)} to study the following attractive Hubbard model on quasicrystals,

\begin{align}\label{s21}
\begin{split}
H=&\sum_{\langle i,j\rangle,\sigma}(t^{\xi_i\eta_j}_{ij}c_{i,\sigma}^\dagger c_{j,\sigma}+{\rm H.c.} )-U\sum_{i}c_{i,\downarrow}^\dagger c_{i,\uparrow}^\dagger c_{i,\uparrow} c_{i,\downarrow}-\mu\sum_{i,\sigma} c_{i,\sigma}^\dagger c_{i,\sigma}.
\end{split}
\end{align}
We have decoupled the interaction term in both the pairing and density channels,
\begin{align}\label{}
\begin{split}
Uc_{i,\downarrow}^\dagger c_{i,\uparrow}^\dagger c_{i,\uparrow} c_{i,\downarrow}\approx c_{i,\downarrow}^\dagger c_{i,\uparrow}^\dagger\Delta_i+\Delta_ic_{i,\uparrow} c_{i,\downarrow}+U\rho_{i,\uparrow} c^\dagger_{i,\downarrow}c_{i,\downarrow}+U\rho_{i,\downarrow}c^\dagger_{i,\uparrow}c_{i,\uparrow}-U\rho_{i,\uparrow}\rho_{i,\downarrow}-{|\Delta_i|^2}/{U},
\end{split}
\end{align}
where $\Delta_i=U\langle c_{i,\uparrow} c_{i,\downarrow}\rangle=U\tilde{\Delta}_i$ and $\rho_{i,\sigma}=\langle c^\dagger_{i,\sigma}c_{i,\sigma} \rangle$ is the site-dependent pairing order parameter and $\rho_{i\sigma}=\langle c^\dagger_{i,\sigma}c_{i,\sigma} \rangle$ is the local density order parameter, respectively.

Under the basis $\Psi = (\hat{c}_{1,\uparrow},\cdots,\hat{c}_{N_t,\uparrow}, \hat{c}_{1,\downarrow}^\dagger, \cdots  \hat{c}_{N_t,\downarrow}^\dagger)^T$, 
where $N_t$ is the total number of sites, the mean-field Hamiltonian becomes
\begin{align}\label{}
\begin{split}
H_{MF}=\Psi^\dagger H_{\rm BdG}\Psi +H_c,
\end{split}
\end{align}
with
\begin{align}\label{s25}
H_{\rm BdG}(\boldsymbol{\phi})=\left[\begin{array}{cc}
h_{\uparrow}(\boldsymbol{\phi})-\mu I-\rho_{\uparrow} U & \Delta \\
\Delta^{*} & -\left(h_{\downarrow}^{T}(\boldsymbol{\phi})-\mu I-\rho_{\downarrow} U\right)
\end{array}\right],
\end{align}
and the constant $H_c=\sum_{i\alpha}\left(U\rho_{i,\uparrow}\rho_{i,\downarrow} +\frac{|\Delta_i|^2}{U}-U\rho_{i,\downarrow}-\mu \right)$. 

In the above BdG Hamiltonian, $h_\sigma(\boldsymbol{\phi}) = \mathcal{H}^0_{\rm rs}(\boldsymbol{\phi})$ denotes the non-interacting part of the Hamiltonian in Eq.~(\ref{s21}). $\Delta$ is the matrix associated with the pairing order parameters, and $\rho_\sigma$ is the matrix associated with the density order parameters, given by
\begin{align}\label{}
\begin{split}
\Delta = \, \text{diag}(\Delta_{1},\cdots,\Delta_{N_t}),\ \ \rho_\sigma=\text{diag}(\rho_{1,\sigma}, \cdots,\rho_{N_t,\sigma}),
\end{split}
\end{align}
respectively. To compute the superfluid weight, we impose closed CBCs on the system, forming a ring in the 1D case and a torus in the 2D case. The magnetic flux parameter \(\boldsymbol{\phi} \) (or equivalentlya uniform vector potential) is then introduced through the Peierls substitution $t_{ij}^{\xi_i\eta_j}\xrightarrow{}t_{ij}^{\xi_i\eta_j}e^{i\boldsymbol{\phi}\cdot\boldsymbol{r}_{i,j}}$, where $\boldsymbol{r}_{ij}=\boldsymbol{r}_j-\boldsymbol{r_i}$ denotes the bond vector pointing from site $i$ to site $j$.

The order parameters are determined self-consistently by minimizing the ground-state energy $E_{\rm GS}(\boldsymbol{\phi})$ of the system. By diagonalizing the BdG Hamiltonian, we obtain the ground-state energy of the system as
\begin{align}\label{}
\begin{split}
E_{\rm GS}(\boldsymbol{\phi})=\sum_{<0}\lambda_i+\sum_{i}\left(U\rho_{i,\uparrow}\rho_{i,\downarrow} +\frac{|\Delta_i|^2}{U}-U\rho_{i,\downarrow}-\mu \right),
\end{split}
\end{align}
where $\lambda_i$ denotes the superconducting quasiparticle eigenvalues of $H_{\mathrm{BdG}}$, and the notation $<0$ indicates that the summation is taken only over the negative quasiparticle eigenvalues.
Then, the superfluid weight can be directly calculated as $D_{s,\alpha\beta}=\frac{1}{L}\frac{d^2F}{d\phi_\alpha d\phi_\beta}$, with the free energy $F=E_{\rm GS}+\mu N$ ($N$ is the total number of particles, which depends on the filling).

When focusing on the half-filling case of a bipartite lattice, the density order parameter becomes a uniform constant~\cite{Uniformdensitytheorem}, i.e., $\rho_{i,\uparrow}=\rho_{i,\downarrow}=\frac12$, and thus does not contribute to the superfluid weight. Therefore, it can be neglected, and the interaction can be decoupled solely in the pairing channel,
\begin{align}\label{}
\begin{split}
Uc_{i,\downarrow}^\dagger c_{i,\uparrow}^\dagger c_{i,\uparrow} c_{i,\downarrow}\approx c_{i,\downarrow}^\dagger c_{i,\uparrow}^\dagger\Delta_i+\Delta_i^{*}c_{i,\uparrow} c_{i,\downarrow}-\frac{|\Delta_i|^2}{U}.
\end{split}
\end{align}
The resulting BdG mean-field Hamiltonian can be obtained by removing the terms involving the density order parameter \( \rho_{i\sigma} \) from the full BdG mean-field Hamiltonian in Eq.~(\ref{s25}),
\begin{align}\label{}
\begin{split}
H^S_{\rm MF}(\boldsymbol{\phi})=\Psi^\dagger H^S_{\rm BdG}(\boldsymbol{\phi})\Psi +\sum_{i,}\left( \frac{|\Delta_i|^2}{U}-\mu \right),
\end{split}
\end{align}
with
\begin{align}\label{}
H^S_{\rm BdG}(\boldsymbol{\phi})=\left[\begin{array}{cc}
h_{\uparrow}(\boldsymbol{\phi})-\mu I & \Delta \\
\Delta^{*} & -\left(h_{\downarrow}^{T}(\boldsymbol{\phi})-\mu I\right)
\end{array}\right].
\end{align}
In the following, we focus exclusively on the half-filling case (the filling $\nu=\sum_{i\alpha,\sigma}\rho_{i\sigma}/(2N_t)=1/2$, and $N_t$ is the total number of sites). Therefore, we can focus solely on the decoupling in the pairing channel. At this point, the ground-state energy is given by
\begin{align}\label{}
\begin{split}
E_{\rm GS}(\boldsymbol{\phi})=\sum_{<0}\lambda_i+\sum_{i}\left(\frac{|\Delta_i|^2}{U}-\mu \right),
\end{split}
\end{align}
where $\lambda_i$ denotes the superconducting quasiparticle eigenvalues of $H^S_{\mathrm{BdG}}$, and the notation $<0$ indicates that the summation is taken only over the negative quasiparticle eigenvalues.

\section{The projection method}
Here, we introduce an analytical projection method that is applicable in the small \( U \) limit. We start from the BdG mean-field Hamiltonian
\begin{align}\label{}
\begin{split}
H_{\rm MF}(\boldsymbol{\phi})=\Psi^\dagger H_{\rm BdG}\Psi +\sum_{i}\left(U\rho_{i,\uparrow}\rho_{i,\downarrow} +\frac{|\Delta_i|^2}{U}-U\rho_{i,\downarrow}-\mu \right),
\end{split}
\end{align}
with
\begin{align}\label{}
H_{\rm BdG}(\boldsymbol{\phi})=\left[\begin{array}{cc}
h_{\uparrow}(\boldsymbol{\phi})-\mu I-\rho_{\uparrow} U & \Delta \\
\Delta^{*} & -\left(h_{\downarrow}^{T}(\boldsymbol{\phi})-\mu I-\rho_{\downarrow} U\right)
\end{array}\right].
\end{align}
Since we focus on the half-filling case, we can adopt the following simplified form, 
\begin{align}\label{}
\begin{split}
H^S_{\rm MF}(\boldsymbol{\phi})=\Psi^\dagger H^S_{\rm BdG}\Psi +\sum_{i,\alpha}\left( \frac{|\Delta_i^\alpha|^2}{U}-\mu \right).
\end{split}
\end{align}

By diagonalizing  $h_{\uparrow}(\boldsymbol{\phi})=h_{\downarrow}(\boldsymbol{\phi})$, we can obtain the eigenspectrum  $\varepsilon(\boldsymbol{\phi})$ and eigenvector matrix $P(\boldsymbol{\phi})$ of the non-interacting Hamiltonian. Then, we can express $H^S_{\rm BdG}$ in the diagonal basis through a transformation $T$,
\begin{align}\label{}
T=\left[\begin{array}{cc}
P(\boldsymbol{\phi}) & 0 \\
0 & P^*(\boldsymbol{\phi})
\end{array}\right].
\end{align}
The BdG Hamiltonian becomes
\begin{align}\label{}
T^\dagger H^S_{\rm BdG}T=\left[\begin{array}{cc}
\varepsilon(\boldsymbol{\phi})-\mu I & P^\dagger(\boldsymbol{\phi})\Delta P^*(\boldsymbol{\phi}) \\
P^T(\boldsymbol{\phi})\Delta^*P(\boldsymbol{\phi}) & -(\varepsilon(\boldsymbol{\phi})-\mu I)
\end{array}\right],
\end{align}
where $\varepsilon(\boldsymbol{\phi})$ represents the diagonal eigenspectrum matrix
\begin{align}\label{eqEnsector}
\varepsilon(\boldsymbol{\phi})={\rm diag}([\varepsilon^1_-(\boldsymbol{\phi}),\cdots, \varepsilon^{N_-}_-(\boldsymbol{\phi}),\varepsilon^1_{\rm f}(\boldsymbol{\phi}),\cdots, \varepsilon^{N_{\rm f}}_{\rm f}(\boldsymbol{\phi}),\varepsilon^1_+(\boldsymbol{\phi}),\cdots, \varepsilon^{N_+}_+(\boldsymbol{\phi})]),
\end{align}
with the superscripts $-, {\rm f}, +$ denoting the eigenvalues of the negative-, flat-, and positive-energy sectors, respectively. \( N_- \), \( N_{\rm f} \), and \( N_+ \) denote the numbers of eigenvalues corresponding to the three respective sectors.

Next, we use a transformation $S$ to separate the contribution from different sectors:
\begin{align}\label{s37}
\begin{split}
ST^\dagger H^S_{\rm BdG}TS^{-1}=S\left[\begin{array}{cc}
\varepsilon(\boldsymbol{\phi})-\mu I & P^\dagger(\boldsymbol{\phi})\Delta P^*(\phi) \\
P^T(\boldsymbol{\phi})\Delta^*P(\boldsymbol{\phi}) & -(\varepsilon(\boldsymbol{\phi})-\mu I)
\end{array}\right]S^{-1}\approx \left[\begin{array}{cc}
\varepsilon_-(\boldsymbol{\phi})-\mu I & P_-^\dagger(\boldsymbol{\phi})\Delta P_-^*(\boldsymbol{\phi}) \\
P_-^T(\boldsymbol{\phi})\Delta^*P_-(\boldsymbol{\phi}) & -(\varepsilon_-(\boldsymbol{\phi})-\mu I)
\end{array}\right]& \\
\oplus \left[\begin{array}{cc}
\varepsilon_{\rm f}(\boldsymbol{\phi})-\mu I & P_{\rm f}^\dagger(\boldsymbol{\phi})\Delta P_{\rm f}^*(\boldsymbol{\phi}) \\
P_{\rm f}^T(\boldsymbol{\phi})\Delta^*P_{\rm f}(\boldsymbol{\phi}) & -(\varepsilon_{\rm f}(\boldsymbol{\phi})-\mu I)
\end{array}\right]
\oplus\left[\begin{array}{cc}
\varepsilon_+(\boldsymbol{\phi})-\mu I & P_+^\dagger(\boldsymbol{\phi})\Delta P_+^*(\boldsymbol{\phi}) \\
P_+^T(\boldsymbol{\phi})\Delta^*P_+(\boldsymbol{\phi}) & -(\varepsilon_+(\boldsymbol{\phi})-\mu I)
\end{array}\right]&,
\end{split}
\end{align}
with the matrix $S$
\begin{align}\label{}
S=\left[\begin{array}{cccccc}
I_{N_-\times N_-} & 0 & 0 & 0 & 0 & 0\\
0 & 0 & 0 & I_{N_-\times N_-} & 0 & 0\\
0 & I_{N_{\rm f}\times N_{\rm f}} & 0 & 0 & 0 & 0\\
0 & 0 & 0 & 0 & I_{N_{\rm f}\times N_{\rm f}} & 0\\
0 & 0 & I_{N_+\times N_+} & 0 & 0 & 0\\
0 & 0 & 0 & 0 & 0 & I_{N_+\times N_+}
\end{array}\right].
\end{align}
Here $P_{s=\pm,{\rm f}}(\phi)=(v_1,v_2,\cdots,v_{N_s})$ is a matrix
of eigenvectors corresponding to the selected sector of eigenstates of non-interacting spectrum, under a magnetic flux $\boldsymbol{\phi}$, and $v_i$ denotes the eigenvector corresponding to the $i$-th eigenstate in $s$ sector. If the structure of the non-interacting energy spectrum of the quasicrystal differs from that in Eq.~(\ref{eqEnsector}), one only needs to appropriately modify the transformation matrix $S$ to similarly separate the different eigenstate sectors.

We retain only the diagonal blocks in Eq.~(\ref{s37}), as the hybridization between different sectors is negligible at small $U$. Moreover, when \( U \) is small, the contributions from the positive- and negative-energy sectors are negligibly small; therefore, only the contribution from the flat-energy sector needs to be retained,
\begin{align}\label{s39}
H_{\rm eff}(\boldsymbol{\phi})\approx \left[\begin{array}{cc}
\varepsilon_{\rm f}(\boldsymbol{\phi})-\mu I & P_{\rm f}^\dagger(\boldsymbol{\phi})\Delta P_{\rm f}^*(\boldsymbol{\phi}) \\
P_{\rm f}^T(\boldsymbol{\phi})\Delta^*P_{\rm f}(\boldsymbol{\phi}) & -(\varepsilon_{\rm f}(\boldsymbol{\phi})-\mu I)
\end{array}\right].
\end{align}
By diagonalizing $H_{\rm eff}(\phi)$, we can obtain all the eigenvalues and, subsequently, the ground-state energy,
\begin{align}\label{}
\begin{split}
E^{\rm proj}_{\rm GS}(\boldsymbol{\phi})=\sum_{<0}\lambda^{\rm f}_i(\boldsymbol{\phi})+\sum_{i\alpha}\frac{|\Delta^\alpha_i|^2}{U},
\end{split}
\end{align}
where $\lambda^{\rm f}_i(\boldsymbol{\phi})$ represent the eigenenergies of the superconducting quasiparticles originating from the flat zero-energy sector, respectively; The label $<0$ indicates that the sum is taken over negative eigenvalues. At half-filling and in the weak \( U \) limit, we can derive the analytical expression for \( \lambda_i^{\text{f}}(\boldsymbol{\phi}) \). Since the Fermi energy lies near the flat zero-energy, we write the chemical potential as $\mu=\varepsilon_{\rm f}(\boldsymbol{\phi})+pU$, where $\varepsilon_{\rm f}(\boldsymbol{\phi})=0$. Then the eigenvalues \( \lambda_i^{\text{f}}(\boldsymbol{\phi}) \) of the second matrix on the right-hand side of Eq. (\ref{s39}) can be obtained as
\begin{align}\label{s311}
\begin{split}
\lambda_i^{\rm f}(\boldsymbol{\phi})=\pm U\sqrt{p^2+d_i(\boldsymbol{\phi})}=\pm U\epsilon_i^{\rm f},
\end{split}
\end{align}
where \( d_i(\boldsymbol{\phi}) \) is the \( i \)-th eigenvalue of the matrix $M^{\rm tot}(\boldsymbol{\phi})$ with
\begin{align}\label{}
\begin{split}
M^{\rm tot}(\boldsymbol{\phi})=P_0^T(\boldsymbol{\phi})\tilde{\Delta}^*P_0(\boldsymbol{\phi})P_0^\dagger(\boldsymbol{\phi})\tilde{\Delta}P_0^*(\boldsymbol{\boldsymbol{\phi}})=P_0^\dagger(-\boldsymbol{\phi})\tilde{\Delta}^*P_0(\boldsymbol{\phi})P_0^\dagger(\boldsymbol{\phi})\tilde{\Delta}P_0(-\boldsymbol{\boldsymbol{\phi}}).
\end{split}
\end{align}
Here, we have defined $\tilde{\Delta}=\text{diag}(\tilde{\Delta}_{1}, \tilde{\Delta}_2,\cdots,\tilde{\Delta}_{N_t})$, with $\ \tilde{\Delta}_i=\Delta_i/U=\langle c_{i,\uparrow} c_{i,\downarrow} \rangle$, and also using the relation $P^*(\boldsymbol{\phi}) = P(-\boldsymbol{\phi})$. 

The superfluid weight, obtained by using the projection method, can be expressed as 
\begin{align}\label{eqdps}
\begin{split}
D_{ps,\alpha\beta}=\left.-\frac{U}{V}\sum_{i}\frac{1}{2\epsilon_i^{\rm f}}\frac{d^2d_i(\boldsymbol{\phi})}{d\phi_\alpha d\phi_\beta}\right|_{\phi_{\alpha,\beta}=0}+\frac{U}{V}C_{\alpha\beta},
\end{split}
\end{align}
where $C_{\alpha\beta}=\sum_i\frac{d^2|\tilde{\Delta}_i|^2}{d\phi_\alpha d\phi_\beta}|_{\phi_{\alpha,\beta}=0}$. Terms that are equal to zero and do not contribute to $D_{ps,\alpha\beta}$ have been omitted. We note that the diagonal elements of the matrix \( \tilde{\Delta} \) are non-uniform, and thus it can be decomposed into a uniform average part $\tilde{\Delta}_0I$ and a fluctuation part $\tilde{\Delta}-\tilde{\Delta}_0I$ with $\tilde{\Delta}_0=\frac{1}{N_t}\sum_{i}\tilde{\Delta}_i(\boldsymbol{\phi}=0)$. 
Consequently, \( M^{\rm tot}(\boldsymbol{\phi}) \) can be decomposed into a part \( M^{\text{uni}}(\boldsymbol{\phi}) \) that contains only \( \tilde{\Delta}_0 \), and a remaining part \( M^{\text{fluc}}(\boldsymbol{\phi}) \) that includes the fluctuations of the pairing order parameters,
\begin{align}\label{}
\begin{split}
 M^{\rm tot}(\boldsymbol{\phi})&=|\tilde{\Delta}_0|^2M^{\text{uni}}(\boldsymbol{\phi})+M^{\text{fluc}}(\boldsymbol{\phi}),\\
 M^{\rm uni}(\boldsymbol{\phi}) &=P_0^\dagger(-\boldsymbol{\phi})P_0(\boldsymbol{\phi})P_0^{\dagger}(\boldsymbol{\phi})P_0(-\boldsymbol{\phi}),\\ \  M^{\rm fluc}(\boldsymbol{\phi}) &=P_0^\dagger(-\boldsymbol{\phi})\tilde{\Delta}^*P_0(\boldsymbol{\phi})P_0^{\dagger}(\boldsymbol{\phi})\tilde{\Delta}P_0(-\boldsymbol{\phi})-|\tilde{\Delta}_0|^2M^{\rm uni}(\boldsymbol{\phi}).
\end{split}
\end{align}
 Since \( [M^{\text{uni}},M^{\text{fluc}}]\approx 0 \), there exists a common set of eigenvectors that diagonalizes them simultaneously.
 Therefore, the first term on the right-hand side in Eq.~(\ref{eqdps}), denoted as \( D_{ps,\alpha\beta}^{(1)} \), can be decomposed into $D^{(1)}_{ps,\alpha\beta}\approx D^{\rm qm}_{ps,\alpha\beta}+D^{\rm fluc}_{ps,\alpha\beta}$, with 
\begin{align}\label{eqdps2}
\begin{split}
D^{\rm qm}_{ps,\alpha\beta}=\left.\frac{U}{V}\sum_{i}\frac{|\tilde{\Delta}_0|^2}{2\epsilon_i^{\rm f}}\frac{d^2d^{\rm uni}_i(\boldsymbol{\phi})}{d\phi_\alpha d\phi_\beta}\right|_{\phi_{\alpha,\beta}=0},\ D^{\rm fluc}_{ps,\alpha\beta}=-\left.\frac{U}{V}\sum_{i}\frac{1}{2\epsilon_i^{\rm f}}\frac{d^2d^{\rm fluc}_i(\boldsymbol{\phi})}{d\phi_\alpha d\phi_\beta}\right|_{\phi_{\alpha,\beta}=0},
\end{split}
\end{align}
where $d_i^{\rm uni}(\boldsymbol{\phi})$ and $d_i^{\rm fluc}(\boldsymbol{\phi})$ are the $i$-th eigenvalues of the matrices $-M^{\rm uni}(\boldsymbol{\phi})$ and $M^{\rm fluc}(\boldsymbol{\phi})$, respectively. The minus sign in front of \( M^{\rm uni}(\boldsymbol{\phi}) \) appears because one of the negative signs in the first term of Eq.~(\ref{eqdps2}) has been absorbed into the second derivative of that term.

It is worth noting that, in $D^{\rm qm}_{ps,\alpha\beta}$, $d_i^{\mathrm{uni}}(\boldsymbol{\phi})$ can be regarded as the $i$-th eigenvalue of the matrix $M_{\rm f}^{\rm rs}(\boldsymbol{\phi}) = I - M^{\mathrm{uni}}$, since the derivative of the identity matrix is zero. We can now see that $M_{\rm f}^{\rm rs}(\boldsymbol{\phi})  = I - P_0^\dagger(-\boldsymbol{\phi})P_0(\boldsymbol{\phi})P_0^{\dagger}(\boldsymbol{\phi})P_0(-\boldsymbol{\phi})$ has the same physical meaning as the quantum distance matrix defined in Eq.~(\ref{eqMrs}), representing the quantum distance between $P(-\boldsymbol{\phi})$ and $P(\boldsymbol{\phi})$. Accordingly, we can define $g_{{\rm f},\alpha\beta,i}^{\mathrm{FS}} = \frac{1}{4}\frac{d^2}{d\phi_\alpha d\phi_\beta}d^{\rm uni}_i(\boldsymbol{\phi})|_{\phi_{\alpha,\beta}=0}$, which is equivalent to the flux-space QM in Eq.~(\ref{eqgni}). The additional factor of $1/4$ arises from the definition of the quantum distance matrix ($I - M^{\mathrm{uni}}$) as the distance between $P(-\boldsymbol{\phi})$ and $P(\boldsymbol{\phi})$.
Then, $D^{\rm qm}_{ps,\alpha\beta}$ can be expressed as, 
\begin{align}\label{}
\begin{split}
D_{ps,\alpha\beta}^{\rm qm}=\left.\frac{U}{V}\sum_{i}\frac{|\tilde{\Delta}_0|^2}{2\epsilon_i^{\rm f}}\frac{d^2d^{\rm uni}_i(\boldsymbol{\phi})}{d\phi_\alpha d\phi_\beta}\right|_{\phi_{\alpha,\beta}=0}=\frac{U|\tilde{\Delta}_0|^2}{2V}\sum_{i}\frac{4g^{\rm FS}_{{\rm f},\alpha\beta,i}}{\epsilon_i^{\rm f}},
\end{split}
\end{align}
representing the contribution associated with the flux-space QM.
\begin{figure}[hbpt]
  \includegraphics[width=17.8cm]{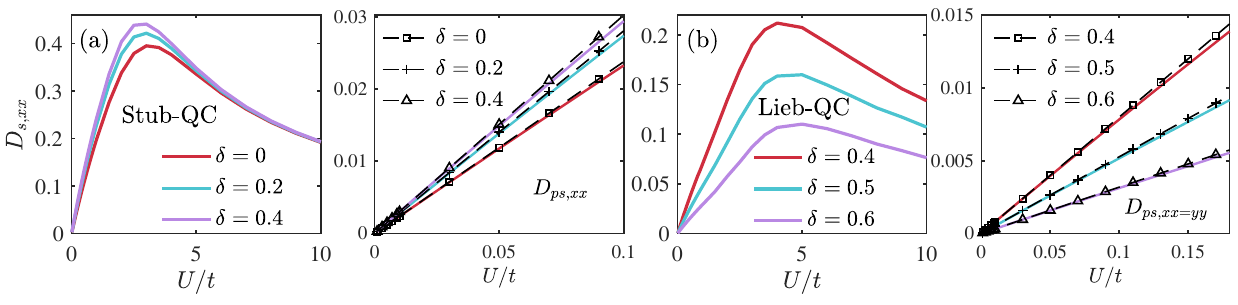}
  \caption{(a) Left panel: the total superfluid weight $D_{s,\alpha\alpha}$ of the Stub-QC as a function of $U/t$; right panel: comparison between $D_{ps,\alpha\alpha}$ and $D_{s,\alpha\alpha}$ in the small-$U$ regime.
(b) Same as in (a), but for the Lieb-QC.
}\label{figSFW}
\end{figure}

\begin{figure}[hbpt]
  \includegraphics[width=12.6cm]{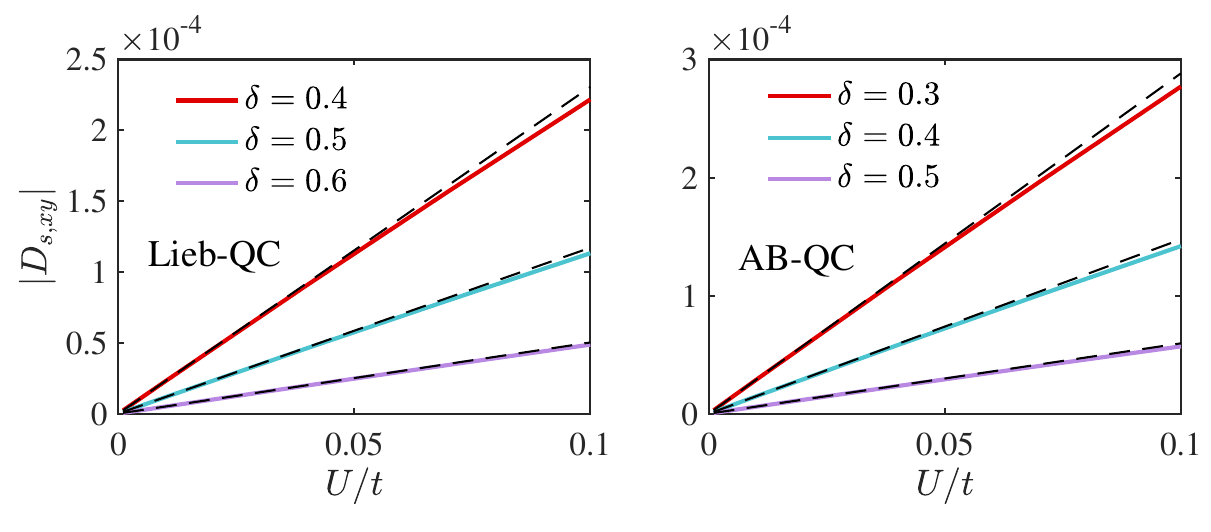}
  \caption{The off-diagonal component of the superfluid weight, \( D_{s,xy} \) (solid lines), for the Lieb-QC (left panel) and AB-QC (right panel) as a function of $U/t$. The dashed lines indicate the corresponding results of \( D_{ps,xy} \) for each value of \( \delta \).}\label{figSDxy}
\end{figure}

Here we present the results of the superfluid weight $D_{s,\alpha\alpha}$ for the Stub-QC and Lieb-QC, as shown in Fig.~\ref{figSFW}. In the small-$U$ regime, $D_{ps,\alpha\alpha} $ in Eq.~(\ref{eqdps}) correctly predicts the linear dependence of the superfluid weight on $U$ as well as its magnitude, revealing the geometric contribution to the superfluid weight originating from the flat-energy states. In Figure.~\ref{figSDxy}, we present the results for the off-diagonal component of the superfluid weight for the Lieb-QC and AB-QC, which is found to be very small. It is worth noting that the overall sign of the off-diagonal term depends on the choice of coordinate orientation; hence, its negative sign can be removed by reversing one of the coordinate axes. Therefore, we show the absolute value of this component.

\begin{figure}[hbpt]
  \includegraphics[width=18cm]{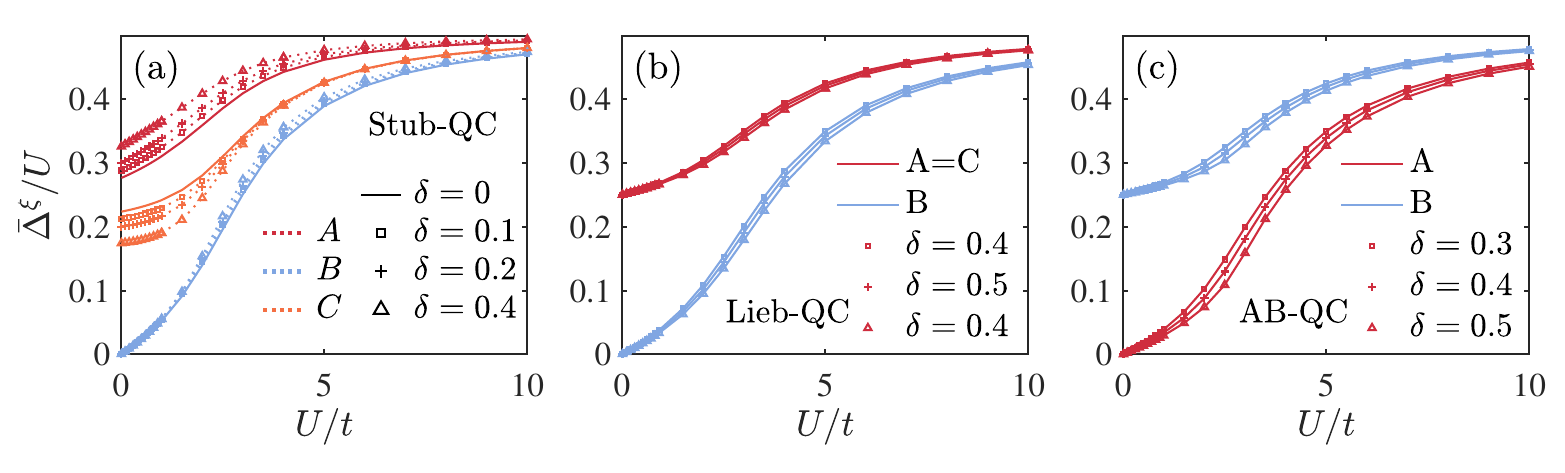}
  \caption{The averaged pairing order parameters on different sublattices, divided by \( U \), as functions of \( U/t \) for (a) Stub-QC, (b) Lieb-QC, and (c) AB-QC.}\label{figSDeltaU}
\end{figure}
In Fig.~\ref{figSDeltaU}, we show the averaged pairing order parameter on each sublattice, defined as \(\bar{\Delta}^\xi = \sum_{i \in \xi} \Delta_i / N_\xi\),
where \(N_\xi\) is the number of sites belonging to sublattice \(\xi\). For the Lieb-QC and AB-QC, \(\bar{\Delta}^\xi / U\) remains nearly identical in the small-\(U\) limit for different values of \(\delta\). In contrast, for the Stub-QC, as \(\delta\) increases, \(\bar{\Delta}^B / U\) remains almost unchanged, while \(\bar{\Delta}^A / U\) increases and \(\bar{\Delta}^C / U\) decreases. Therefore, for all three types of quasicrystals, in the small-\(U\) limit, the average pairing order parameter over all sites divided by \(U\) remains nearly unchanged as \(\delta\) varies. That is, \(\tilde{\Delta}_0 = \frac{1}{U} \sum_i \frac{\Delta_i(\boldsymbol{\phi}=0)}{N_t}\) remains nearly constant as \(\delta\) varies in the small-\(U\) limit.

\section{Wannier function in quasicrystals}
\begin{figure}[hbpt]
  \includegraphics[width=16cm]{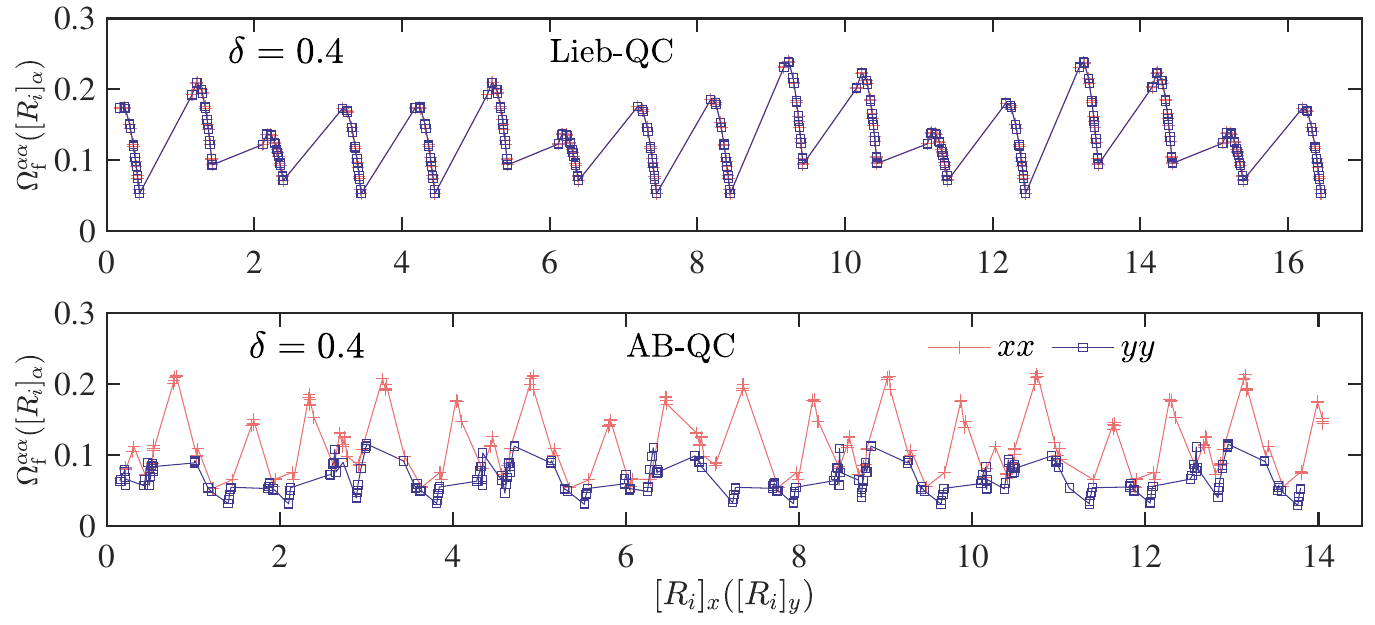}
  \caption{For the parameter $\delta=0.4$, the Wannier function spread functional at different Wannier centers for the Lieb-QC (top panel) and AB-QC (bottom panel). $xx$ and $yy$ denote the results of \( \langle\Omega^{xx}_{s}\rangle_V \) and \( \langle\Omega^{yy}_{s}\rangle_V \), respectively. The horizontal axis represents the centers of the Wannier functions.
}\label{figS_OmegaR}
\end{figure}

\begin{figure}[hbpt]
  \includegraphics[width=12cm]{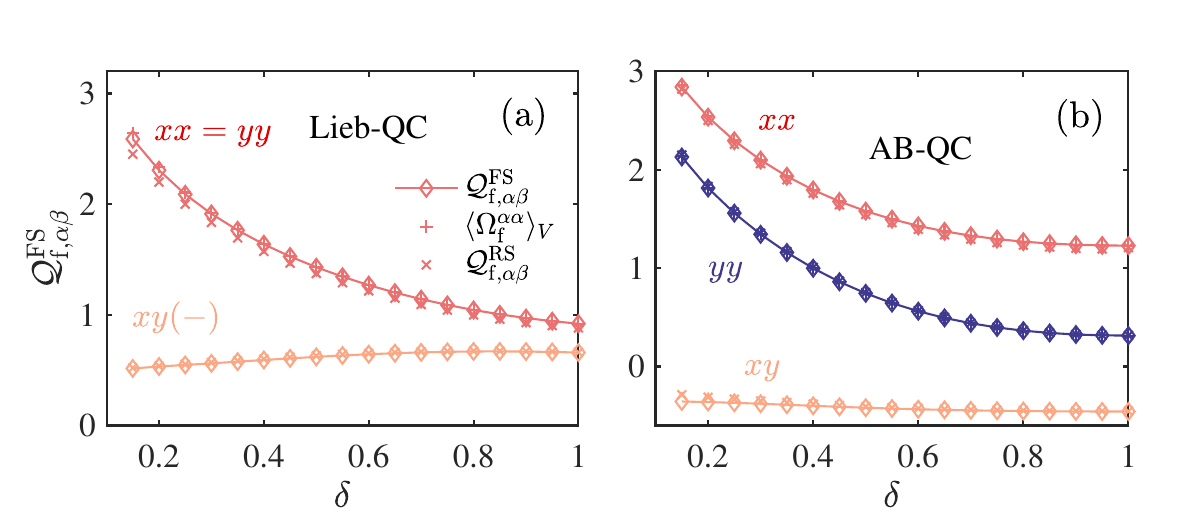}
  \caption{For the Lieb-QC (a) and AB-QC (b), the quantities \(Q^{\rm FS}_{{\rm f},\alpha\beta}\), \( \langle\Omega^{\alpha\alpha}_{{\rm f}}\rangle_V \), and \(Q^{\rm RS}_{{\rm f},\alpha\beta}\) corresponding to the flat-energy sector are shown as functions of \( \delta \).
}\label{figS_IQM}
\end{figure}
For a general \(d\)-dimensional system, the maximally localized Wannier functions $w_s([r-R_i]_\alpha)$ along a given direction \(\alpha\) can be obtained by computing the eigenfunctions of the projected position operator \(\hat{P} r_\alpha \hat{P}\) associated with a selected eigenstates sector $s$~\cite{wfdisordersoliton(1982)}:
\begin{align}\label{eqPxP}
\hat{P}\hat{r}_\alpha\hat{P}|w_s([r-R_i]_\alpha)\rangle=X_i |w_s([r-R_i]_\alpha)\rangle,
\end{align}
where $\hat{P}=\sum_{i\in s}|\varphi_i\rangle\langle\varphi_i|$ denotes the projection operator onto the selected eigenstates sector.
Under open boundary conditions, the position operator $\hat{r}_\alpha=r_\alpha$ corresponds to the $\alpha$-component of the lattice site coordinates, and the eigenvalues of projected position operator represent the localization centers of the Wannier functions along the $\alpha$ direction: $[R_i]_\alpha=X_i$. Under closed boundary conditions (CBC), however, the position operator must be replaced by $\hat{r}_\alpha= e^{i2\pi r_\alpha/L_\alpha}$~\cite{Resta1998PRL}, in which case the center of the Wannier function corresponds to $[R_i]_\alpha=\frac{L_\alpha}{2\pi}{\rm Im}(\ln X_i)\ {\rm mod}\  L_\alpha$. 

For each maximally localized Wannier function along a direction \(\alpha\), $|w_s([r-R_i]_\alpha)\rangle$, the spread functional can be computed as
\begin{align}\label{s53}
\begin{split}
\Omega^{\alpha\alpha}_s([R_i]_\alpha)&=\langle {r}_\alpha^2 \rangle_s-\langle {r}_\alpha\rangle_s^2=\langle([r-R_i]_\alpha)^2\rangle_s=\sum_r ([r-R_i]_\alpha)^2|w_n([r-R_i]_\alpha)|^2,
\end{split}
\end{align}
where the average $\langle\cdot\rangle_{\rm s}$ is taken over the Wannier function \(w_s([{r} - {R}_i]_\alpha)\). For the flat-energy states sector ($s={\rm f}$), we find that the IFSQM is equivalent to the volume-averaged spread functional, denote as $\langle\Omega^{\alpha\alpha}_{\rm f}\rangle_V= 2\frac{(2\pi)^{n-1}}{V}\sum_{R_i}\Omega^{\alpha\alpha}_{\rm f}([R_i]_\alpha)$, i.e., \(Q^{\rm FS}_{\rm f,\alpha\alpha} = \langle\Omega^{\alpha\alpha}_{\rm f}\rangle_V\). Thereby establishing a direct relation between the IFSQM and the Wannier function.

 Moreover, our IFSQM is equivalent to the real-space integrated quantum metric (RSIQM) $\mathcal{Q}^{\rm RS}_{s,\alpha\beta}$ reported in previous works~\cite{Resta2018,PhysRevLett.122.166602,PhysRevB.107.205133,PhysRevB.111.134201}, 
\begin{align}\label{eqRSIQM}
\begin{split}
 \mathcal{Q}^{\rm RS}_{s,\alpha\beta}=-\frac{(2\pi)^{n-1}}{V}{\rm Re}\ {\rm Tr}\left\{\hat{P}\left[\hat{r}_\alpha,\hat{P}\right]\left[\hat{r}_\beta,\hat{P}\right]\right\},
\end{split}
\end{align}
where $\hat{P}=\sum_{j\in s}|\varphi_j\rangle\langle\varphi_j|$ is the projection operator onto the \( s \)-sector. It is worth noting that the method shown in Eq.~(\ref{eqRSIQM}) is computed under open boundary conditions, and therefore requires a larger system size to obtain converged results. In Fig.~\ref{figS_IQM}, we demonstrate the equivalence between the IFSQM and RSIQM, i.e., \(Q^{\rm FS}_{s,\alpha\beta} =\mathcal{Q}^{\rm RS}_{s,\alpha\beta}\), as well as the equivalence between their diagonal components and the volume-averaged Wannier function spread, i.e., \(Q^{\rm FS}_{s,\alpha\alpha} = \langle\Omega^{\alpha\alpha}_{s}\rangle_V=\mathcal{Q}^{\rm RS}_{s,\alpha\alpha}\).

\bibliographystyle{apsrev4-1-etal-title_10authors}
\bibliography{ref}
